\RequirePackage{snapshot}
\documentclass[APA,Times2COL]{WileyNJDv5}
\usepackage{silence}  % Allows selective suppression of warnings
\usepackage{comment}
\usepackage{subcaption}
\usepackage{amsmath}
\usepackage{cuted}
\usepackage{tabularx}

\articletype{Research article}
\startpage{1}
\usepackage[T1]{fontenc}      % Use T1 encoding for scalable fonts
\usepackage{lmodern}          % Latin Modern, scalable Computer Modern replacement
\usepackage{eucal}            % Euler script fonts for \mathscr at all sizes
\DeclareMathAlphabet{\mathscr}{U}{eus}{m}{n} % Ensure scalable \mathscr
\usepackage[switch,mathlines]{lineno}
\begin{document}
\title{A phase-field modeling approach to sea-ice fracturing}
\author[1,2]{Laetitia Drumare}
\author[2]{Vidar Skogvoll}
\author[1,2]{Fran\c cois Renard}
\author[2]{Luiza Angheluta}
\author[1,3]{Véronique Dansereau}
\titlemark{Phase-field modeling approach to sea-ice fracturing}
\address[1]{\orgdiv{ISTerre}, \orgname{Univ. Grenoble Alpes, Univ. Savoie Mont Blanc, Univ. Gustave Eiffel, CNRS, IRD}, \orgaddress{\state{Grenoble}, \country{France}}}
\address[2]{\orgdiv{Njord Centre, Department of Physics and Geosciences}, \orgname{ University of Oslo}, \orgaddress{\state{Oslo}, \country{Norway}}}
\address[3]{\orgdiv{IGE}, \orgname{Univ. Grenoble Alpes, Univ. Savoie Mont Blanc, CNRS, IRD}, \orgaddress{\state{Grenoble}, \country{France}}}
\corres{Corresponding authors: Laetitia Drumare \email{laetitia.drumare@univ-grenoble-alpes.fr} and Véronique Dansereau \email{veronique.dansereau@univ-grenoble-alpes.fr}.}
\abstract[Abstract]{
The thin ice that covers the polar oceans is a complex geomaterial that is constantly stressed and fractured by winds and ocean currents. In the central Arctic, this forcing produces deformations in the form of shear bands, within which individual ice plates detach, locally generating a granular medium. Capturing this transition from a continuous to a granular sea-ice cover has implications for the adequate representation of the mechanical and dynamical behavior of sea-ice in regional and large-scale models used for operational and climate prediction purposes. Our work investigates the feasibility of a phase-field approach to capture this granularization processes and focuses on fracture propagation in the material. The model combines a double-well free-energy formulation with an overdamped displacement response. The governing equations are solved using a spectral method in Fourier space. The implementation accounts for body forces, representative of the main forcings on sea-ice, and remains computationally tractable despite the highly nonlinear character of the double-well energy formulation. We first validate the framework against a benchmark problem: the opening of an inclusion embedded into an elastic matrix under tensile loading. Then, additional simple shear configurations are investigated: an inclusion solicited under plane shear and a cylindrical Couette experiment, for which the analytical solution of the displacement field is known. The resulting fracture patterns and displacement fields demonstrate that our phase-field framework captures key features of tensile and shear fracture propagation, including the linear scaling between crack speed and applied load predicted by the Griffith's theory.
}
\maketitle
% \linenumbers
%
\section{Introduction}\label{sec1}
Fracturing plays a fundamental role in the deformation of many geophysical systems, including tectonic faults, landslides, glaciers, and sea-ice. Among these, sea-ice exhibits some of the fastest dynamics and is distinctive in being active across all timescales from seconds, days, and seasons \cite[e.g.,]{Herman2012, Oikkonen2017, Weiss2017, uusinoka2025, Liu2025}. Indeed, this thin layer of floating ice that covers the polar oceans fractures constantly under the stresses imposed by the winds, ocean currents and waves. Once broken, it can drift at speeds as fast as several tens of kilometers per day \cite[e.g.,]{lavergne2023}. This brittle behavior controls both its local and large-scale dynamics by localizing deformation into narrow bands with high fracture density known as Linear Kinematic Features \cite{kwok2001}, which span spatial scales from meters to hundreds of kilometers. Fractures also generate openings that directly connect the ocean and atmosphere, thereby strongly modulating exchanges of heat, momentum, gases, and biogeochemical tracers between the underlying ocean and overlying atmosphere \cite{serreze2011, goosse2018, Taylor2018}. Accurate representation of how sea-ice fractures and deforms is therefore critical for reliable modeling of its evolution under a changing climate \cite{ipcc_underestimation}.

To represent the fracturing processes in sea-ice, current continuum models used for operational and climate purposes mostly rely on two different approaches: viscous-plastic rheologies that account for the effect of fracturing on the evolution of sea-ice concentration (describing the surface area covered by ice versus open water, \cite{hibler1979, hunke1997}) and elasto-brittle rheologies \cite{girard2011, dansereau2016, olason2022}, which additionally account for the evolution of a spatially-averaged fracture density through the incorporation of a progressive damage mechanism. While computationally efficient and therefore relevant to large-scale settings, neither of these approaches explicitly capture single crack nucleation, propagation, evolving geometry nor crack interactions. Understanding how individual fractures initiate and propagate under a volumetric, distributed forcing - the wind and ocean drags - and how fractures interact to eventually produce high-density fracture zones, is the main motivation for this work. A solid understanding of these processes would indeed be useful to determine whether the fractures exhibit characteristic geometrical patterns and whether individual ice plate, so-called floes, generated in such shear bands exhibit a characteristic length scale in their size distribution. Importantly, it would also allow quantifying how these localized fracture events modify the mechanical properties of sea-ice at larger scales.

In this context, we explore phase-field models of fracture, which naturally handle complex geometries and crack interactions without explicit tracking of crack surfaces. In classical linear elastic fracture mechanics, crack propagation is characterized by stress intensity factors associated with sharp crack tips, leading to singular stress fields and requiring explicit tracking of crack geometry \cite{Irwin1957,anderson2005}. Alternatively, phase-field models approximate sharp cracks by a diffuse field, thereby eliminating crack-tip singularities and providing a computationally robust alternative to discrete crack.

This diffuse field is defined through an order parameter that locally describes the degree of brokenness of the material. As a result, phase-field approaches are able to capture both fracture evolution in a localized region around the crack tip (via strain accumulation) together with the macroscopic linear elastic behavior in the bulk material. The main objective of this work is to explore the feasibility of the phase-field approach as a mechanical framework for sea-ice modeling. Based on the simple framework presented below, our future work will address the increased complexity of sea-ice fracturing.
\begin{figure*}[htbp]
    \centering
    \begin{subfigure}[t]{\textwidth}
        \centering
        \includegraphics[width=0.35\linewidth]{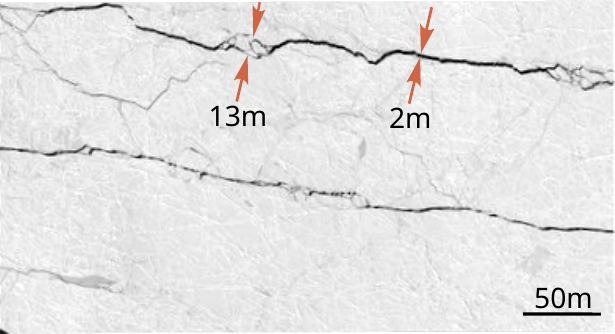}
        \put(0,0){\textbf{(a)}}
        \phantomcaption
        \label{fig:seaice_shear_band_early_stage}
    \end{subfigure}
    \hfill
    \begin{subfigure}[t]{0.46\textwidth}
        \centering
        \includegraphics[width=0.77\linewidth]{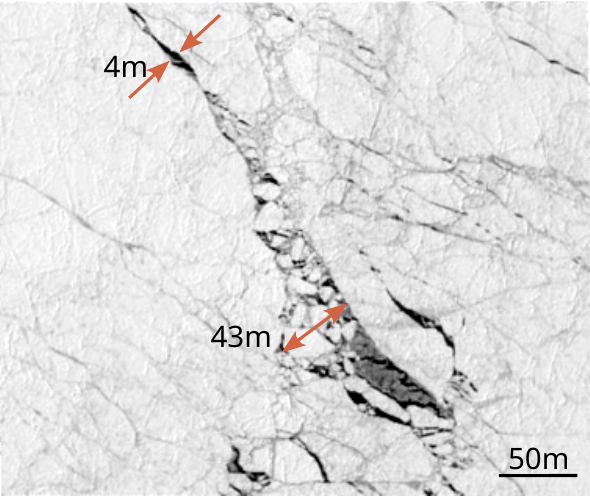}
        \put(0,0){\textbf{(b)}}
        \phantomcaption
        \label{fig:seaice_shear_band_meters}
    \end{subfigure}
    \hfill
    \begin{subfigure}[t]{0.46\textwidth}
        \centering
        \includegraphics[width=0.9\linewidth]{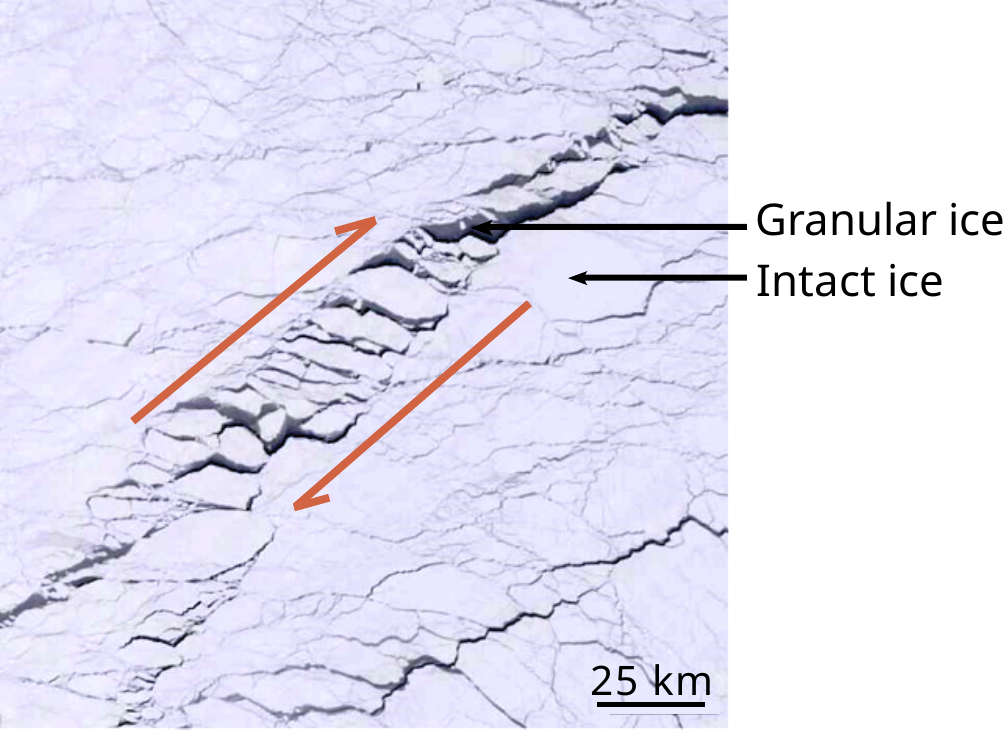}
        \put(0,0){\textbf{(c)}}
        \phantomcaption
        \label{fig:seaice_shear_band_kilometers}
    \end{subfigure}
    \caption{Satellite images of shear fractures in sea-ice at different scales and different deformation stages. (a) Early stage deformation, showing a single dominant fracture prior to gouge formation; (b) resulting granular state of ice within a shear band at small scale. USGS EarthExplorer images courtesy of the U.S Geological Survey, \protect\cite{imageseaice_usgs}. (c) Resulting granular state of ice within a mature shear band at larger scale. Image generated using NASA Worldview (NASA ESDIS), \protect\cite{nasaworldview}.}
    \label{fig:description_of_ice}
\end{figure*}
\section{Background: phase-field fracture}
Several phase-field models for fracture propagation have been formulated based on an underlying energy functional minimization. However, they differ in how this energy governs crack evolution. A key distinction between two main approaches is whether energy minimization is imposed globally or as dissipative relaxation.

The first of these approaches gathers the variational models, which frame the fracture problem as a competition between the elastic strain energy stored in the bulk of the material and the energy required to create new fracture surfaces. In this framework, fracture is interpreted as a geometric discontinuity whose evolution is governed by a global energy balance. Fracture propagation is therefore treated as an energy minimization problem with irreversibility \cite{francfort_marigo_1998}, with the aim to provide a rigorous variational justification of Griffith's fracture theory \cite{griffith1921}. In this model, cracks are still represented as sharp discontinuities. Independently, an elliptic regularization approach was developed in the context of image segmentation and free-discontinuity problems \cite{mumfordshah1989,ambrosio_tortorelli_1990}. It replaces sharp discontinuities by a diffuse interface described by a phase-field of finite width. Within this approach, as the width of the interface tends to zero, the regularized functionals $\Gamma$-converge to the original functional with sharp discontinuities \cite{ambrosio_tortorelli_1990}. Later, this regularization framework was recognized as a suitable phase-field approximation of the brittle fracture energy proposed by~\cite{francfort_marigo_1998}.

Building on this connection, phase-field models of fracture suitable for engineering applications were developed in the early 2000's \cite{bourdin2000, bourdin2007, bourdin2008}. These models couple the regularization with finite elements and incorporate a split of the strain energy to break the symmetry in traction and in compression \cite{amor2009}. This formulation became the standard quasi-static computational phase-field fracture method. The key characteristics of this approach are that the crack is described as an interface embedded within a single stable phase (the intact material), rather than a stable phase of its own; that it is naturally quasi-static, as the underlying theory is about energy minimization rather than dynamics; and that it reproduces Griffith-type fracture within a mathematically rigorous, variational framework.

Another phase-field fracture model motivated by modeling dynamic crack propagation, including branching and instabilities observed in high-speed fracture experiments, was proposed by ~\cite{Karma_Kessler_Levine}, hereafter referred to as the KKL model. Within this framework, fracture is treated as a phase transition between intact and broken states, described by a phase-field governed by a double-well potential and a gradient term for surface energy \cite{colline_levine1985, karma_rappel1998}.

This formulation was shown to accurately capture the expected behavior of mode-III cracks. It was subsequently extended within a two-dimensional plane strain configuration to investigate mode-I and mode-II crack behavior, while retaining the (damped) dynamic formulation \cite{henry_levine2004, karma_lobkovsky2004}. These authors further examined crack tip instabilities, branching phenomena, and the occurrence of supersonic crack propagation. To prevent fracturing under compression, a breaking of the symmetry of the elastic energy between compression and tension was also introduced \cite{henry_levine2004}. This particular formulation allows dynamic crack growth. However, crack irreversibility must be enforced artificially, since the material inherently has two stable states, and, depending on the local strain energy, it can naturally heal. The elastic response is governed by an overdamped equation of the displacement field. The phase-field evolution governing the fracture propagation has its own dynamics. The timescale separation has to be parametrized carefully depending on the material response, that is, brittle or ductile. The applicability of the KKL model to dynamic fracture has thus been investigated, and it has been shown that an extremely simple model grounded in generic physical considerations is capable of reproducing a broad range of crack patterns \cite{henry_levine2004}. The KKL category of models focuses on realistic crack propagation under rapid loading, not strictly on energy convergence. The phenomenology of KKL models is worth being investigated, specifically if one studies the interaction of different cracks, as the phase-field method can easily deal with intersecting interfaces \cite{henry_levine2004}.
\section{Model description}\label{sec2}
\subsection{Model formulation}
Based on the above non‑exclusive review of the two main categories of phase‑field formulations for fracture mechanics, the following section outlines the fracturing events to be modeled and examines the geophysical context to justify the adopted modeling choices.

Sea‑ice fracturing arises under different types of mechanical loading. Tensile loading occurs, for example, in the so-called \textit{land‑fast} ice, where the ice is pulled away from the shore by the winds. It also occurs locally when the ice cover is being bent by waves \cite{thomson2014}. Over the central Arctic, where the ice is dense and shielded from waves, compressive shear driven by winds and ocean currents appears to be a dominant mode of mechanical loading, standing out in \textit{in-situ} stress data \cite[e.g.,]{schulson2004} and deformation analyses \cite{kwok2001}.

Fractures formed under shear loading present a rough, self-affine geometry whose fluctuations obey a scale-invariant power law \cite{weiss2002}, so that irregularities, and the resulting mechanical interlocking and friction, persist over a broad range of scales, from sub-millimetre to kilometre scales \cite{weiss2002}. Under compressive–shear loading, this multiscale roughness promotes secondary fracturing, concentrated grain breakage and sliding, ultimately producing a shear band filled with fragmented ice that behaves as a granular gouge, akin to that observed in tectonic faults \cite{kwok2001, schulson2004}. Figure~\ref{fig:seaice_shear_band_early_stage} shows a fresh, rough fracture at an early stage prior to gouge formation, although localized zones of incipient granularization are already visible. The granular state of ice within the shear band at a more advanced stage is observed in Figure~\ref{fig:seaice_shear_band_meters} and ~\ref{fig:seaice_shear_band_kilometers} at two different scales, thereby illustrating that the same deformation process operates over markedly distinct length scales. In this shear band scenario, two states of sea-ice can be distinguished: the relatively intact ice surrounding the band, and the fragmented ice within (Figure~\ref{fig:seaice_shear_band_kilometers}). These two kinds of ice have different mechanical and dynamical behaviors: the former is a cohesive, elastic, solid - in which further fracturing can occur - the second is a granular material - dominated by collision, friction and healing processes. Considering that these two states coexist at equilibrium within sea-ice agrees with a double-well energy description as in the KKL formulation.
\subsubsection{Relevant length scales}
We can further justify this choice through a dimensional analysis of the system. Within the phase-field fracture framework, the phase-field variable that smoothly transitions from intact to fully fractured material introduces an intrinsic length scale $l_0$, which controls the width of the diffuse interface and serves as a numerical regularization parameter for the phase-field gradient. From a computational perspective, $l_0$ must be chosen sufficiently large to ensure that the interface is properly resolved by the discretization, typically requiring at least five grid points across the diffusive zone. In the present study, with a spatial resolution $\Delta x = \Delta y = 1\mathrm{m}$, this implies $l_0 \gtrsim 5\mathrm{m}$ to accurately capture the phase-field gradient. At the same time, $l_0$ must remain small enough to resolve fractures as thin as possible, as it directly sets the minimum width of damage that can be represented. 
It should be emphasized that $l_0$ is a numerical parameter, not linked with any intrinsic mechanical property of the material. As a result, the model is in principle applicable across a wide range of spatial scales, with the primary constraint being the achievable resolution relative to the domain size. In this work, we do not seek to resolve the smallest possible fractures or the finest roughness features due to computational limitation. Instead, roughness effects are implicitly represented through the coexistence of two distinct phases, corresponding to intact and granular ice.

From a physical perspective, the types of fracturing events considered here occur over a wide range of spatial scales, from tens of meters (Figure~\ref{fig:seaice_shear_band_early_stage} and~\ref{fig:seaice_shear_band_meters}) to several tens of kilometers (Figure~\ref{fig:seaice_shear_band_kilometers}). As the aim is ultimately to model several interacting fractures within the domain, the system size must be large enough to accommodate multiple fractures while remaining small enough to keep the numerical cost reasonable. In the benchmark tests presented in the present study, however, we restrict ourselves to a simpler configuration in which a single main fracture (with two or four branches) propagates from a heterogeneity (inclusion). For these benchmark cases, a domain size of $100$m×$100$m is therefore appropriate, while larger domains will be considered in future studies.

As fractures thinner than the length scale $l_0$ are not resolved, the present model is best interpreted as describing the evolution of a finite-width damage zone rather than discrete, sharply defined, small-scale fractures.
\subsubsection{Relevant time scales}
Considering the timescales of the fracturing phenomenon and according to \cite{weiss2013} and \cite{rampal2009}, the quasi-static assumption is valid for large time scales greater than one day. Under this assumption, inertial forces are much smaller than other forces. However, the time scales of interest for fracturing events range from one second to one hour. Consequently the quasi-static hypothesis is no longer appropriate. We can verify this by using geophysical orders of magnitude to confirm the hypothesis of the present modeling framework. To this end, we consider a simplified form of the momentum equation of sea-ice, following \cite{steele1997}:
\begin{equation}
    \partial_t U_i = F_{\mathrm{air-drag}} + F_{\mathrm{ocean-drag}} + F_{\mathrm{internal-stress}} + F_{\mathrm{Coriolis}} + F_{\mathrm{sea-surface-tilt}}.
    \label{eq:order_magnitude_momentum_eq_seaice}
\end{equation}
%
% \resetlinenumber
% \linenumbers
Here, $\partial_t U_i$ is the local acceleration of the ice. The contributions from the Coriolis force and sea surface tilt are typically small, even more considering the large domain size, and often approximately cancel out each other as they are oppositely directed, allowing them to be neglected in a first-order analysis \cite{steele1997, weiss2013}. Even if linear drag force can be appropriate for ocean drag, a quadratic form of the drag force is preferred for both water ($F_{\mathrm{water}}$) and wind ($F_{\mathrm{air}}$) drag~\cite{weiss2013}:
\[
\begin{aligned}
F_{\mathrm{water}} &= \rho_w C_w(U_w-U_{\mathrm{ice}})^2 \quad \approx \quad 3.10^{-5}\quad \text{Pa},\\
F_{\mathrm{air}} &= \rho_a C_aU_a^2 \quad \approx \quad  5.10^{-2} \quad \text{Pa}.
\label{eq:oom_external_forcing}
\end{aligned}
\]
To estimate the magnitude of the inertial term, we used a decomposition of the sea-ice velocity into a mean field and a fluctuating velocity field noted $U'$, derived from buoy data (International Arctic Buoy Program \cite{iabp, iabp_website}), with a time resolution $\tau$ of three hours \cite{weiss2013, rampal2009}. The inertial contribution can be then estimated as:
\[
\partial_t U_{\mathrm{ice}} = \rho_{\mathrm{ice}} h \frac{\sqrt{(\Delta U')^2}}{\tau} \quad \approx \quad 2.10^{-3} \quad \text{Pa},
\]
where $\sqrt{<(\Delta U')^2>} \approx 5$m/s represents the typical incremental velocity and is calculated by $(\Delta U')^2 = |U'(t+\tau)-U'(t)|^2$.

Finally, the internal stress term can be estimated as $\nabla( \sigma h)$, where $h$ is the average thickness of sea-ice, and $\sigma$, the internal stress, can be approximated from in situ measurements using vibrating wire sensors, which provide access to the two-dimensional stress tensor at a given location. These measurements suggest a typical cohesion (resistance of the material in simple shear) ranging from 30 to 60 kPa \cite{weiss2013}, which can be seen as the maximum stress difference sustained over a length scale of about 100~km. That gives an estimation for the magnitude of the internal stress term at $10^{-1}$~Pa.

In summary, the large-scale motion of sea-ice is primarily governed by the balance between internal stresses and external wind forcing \cite{steele1997, weiss2013}. However, over the time scales considered here, that is seconds to hours, the rapid variability of the wind forcing implies that inertial effects cannot be systematically neglected, so that a \textit{dynamic} description is required.

Building on the multiscale picture involving friction along rough sea-ice fractures discussed above \cite{weiss2002}, the present model represents a small associated energy dissipation through an overdamped evolution of the elastic displacement field under external forcing. The mechanical problem is thus treated dynamically, but with overdamped displacement dynamics, which effectively captures a viscous-like resistance arising from sliding and interlocking along rough fracture boundaries. Accordingly, the KKL-formulation is adopted both for the two stable phases of the material and for the overdamped-dynamical evolution. Note also that in the present study, we are considering only static forces, and no dynamical variations of the loading conditions. Implementing such dynamical effects will be a topic of a future study.
\subsubsection{Loading conditions and numerical feasibility}
Owing to the particular geophysical context of sea-ice, the implementation of mechanical loading has important specific implications. Indeed, in  many fracture problems, such as the modeling of tectonic faults, loading is prescribed as Dirichlet boundary conditions. In the case of sea-ice however, the geophysical forcing corresponds to volumetric body forces. To better reflect this specificity, we therefore apply body forces in our model rather than boundary displacements. Our choice of discretization using a spectral method in Fourier space is consistent both with the body forcing (respecting periodic boundary conditions) and the numerical simplicity and efficiency of the implementation, especially since the double-well potential formulation provides a highly non-linear evolution equation.
\subsection{Model equations}
We consider a two-dimensional isotropic, linear-elastic solid undergoing fracture under external loading. Fracture is described using a phase-field variable $\phi(\mathbf r,t)$, where $\mathbf r$ denotes position and $t$ time. The $\phi$-field varies smoothly between two bulk states, i.e., $\phi=1$ in the intact solid and $\phi=0$ in the fully broken (granular) phase. These values correspond to the minima of a double-well potential,
\[
V_{DW}(\phi) = \frac{1}{4}\phi^2(1-\phi)^2,
\]
following the standard KKL formulation \cite{Karma_Kessler_Levine}. The total free energy is defined as:
\begin{equation}
    E= \int dr^2 \frac{\kappa_1}{2} |\nabla \phi|^2 + \kappa_2 V_{DW}(\phi) + g(\phi) \left( \mathscr{E}_{\text{strain}}-\mathscr{E}_c \right),
    \label{eq:free_energy_dim}
\end{equation}
where $\mathscr{E}_{\text{strain}}$ is the elastic strain energy density (per unit area), $\kappa_1$ controls the interfacial energy cost associated with gradients in $\phi$, and $\kappa_2$ sets the height of the double-well barrier. The parameter $\mathscr{E}_c$ is the critical strain energy density for mode-I fracture, consistent with Griffith’s criterion \cite{griffith1921}, which states that crack growth occurs when elastic energy exceeds the energetic cost of creating new surfaces. The functional in Eq.~\eqref{eq:free_energy_dim} therefore contains three contributions: (i) an interfacial energy penalizing spatial variations of $\phi$, (ii) a local double-well potential enforcing phase separation, and (iii) a coupling between the phase-field and the elastic energy that determines whether the material locally remains intact or transitions to a broken state. This coupling is controlled by a degradation function $g(\phi)$, which must satisfy three conditions: (1) $g(0)=0$, ensuring vanishing elastic energy in the fully broken state; (2) $g(1)=1$, recovering the full elastic energy in the intact phase; and (3) $g'(0)=g'(1)=0$, ensuring that $\phi=0$ and $\phi=1$ remain stationary points of the energy. Following the standard KKL choice, we consider the form
\[
g(\phi)=4\phi^3 - 3\phi^4,
\]
as proposed in Ref.~\cite{Karma_Kessler_Levine, HAKIM2009342, karma_lobkovsky2004, henry_levine2004}.

Consistent with the aspect ratio of sea-ice, which extends over thousands of kilometers while remaining a few meters thick, we consider a thin, two-dimensional horizontal plate under plane stresses. The elastic strain energy density per unit of area  $\mathscr{E}_{\text{strain}}$ is therefore given by:
\begin{equation}
\mathscr{E}_{\text{strain}} = \frac{Yh}{(1+\nu)(1-\nu)} \left( \varepsilon_{xx}^2 + \varepsilon_{yy}^2 + 2 \nu\varepsilon_{xx}\varepsilon_{yy} \right) + 2\frac{Yh}{1+\nu}\varepsilon_{xy}^2,
\end{equation}
where $Yh$ is the in-plane elastic stiffness of the intact material with Young’s modulus $Y$ and average thickness $h$, and $\varepsilon_{ij} = \frac{1}{2}\left(\frac{\partial u_j}{\partial x_i}+\frac{\partial u_i}{\partial x_j} \right)$ is the strain tensor computed from the horizontal displacement field ($u_x,u_y$).

The phase-field follows a gradient descent of the free energy and therefore captures the dissipative evolution of fracture propagation. This evolution is coupled to the momentum balance, describing the dynamic response of the material under loading. We consider an overdamped regime in which frictional damping relaxes the displacement field to mechanical equilibrium in the presence of external loading, while the main source of dissipation is the fracturing process itself. The governing equations are given by Eq. \ref{eq:evolution_equation_PF}.
\begin{figure*}[h!]
    \begin{equation}
        \begin{cases}
            \partial_t\phi &= - \chi \frac{\delta E}{\delta \phi} = \chi\left (\kappa_1 \nabla ^2 \phi - \kappa_2 V'_{DW}(\phi) - g'(\phi)(\mathscr{E}_{\text{strain}}-\mathscr{E}_c) \right) \\
            \rho \beta\partial_t u_i &= \partial_j \sigma_{ij} + f_i^{ext}\\
            \mathscr{E}_{\text{strain}} &= \frac{Yh}{(1+\nu)(1-\nu)} \Big[ (\partial_x u_{x})^{2} + (\partial_y u_{y})^{2} + 2\nu\,\partial_x u_{x}\partial_y u_{y} \Big] + \frac{Yh}{2(1+\nu)} \big(\partial_x u_{y} + \partial_y u_{x}\big)^{2}
        \end{cases}.
    \label{eq:evolution_equation_PF}
    \end{equation}
\end{figure*}
In the phase-field equation, $\chi$ is a kinetic coefficient controlling the rate of energy dissipation in the fracture process zone and ensuring correct dimensional consistency \cite{HAKIM2009342, allencahn1972}. External loading is applied either through a prescribed body force $f_i^{ext}$ or via boundary-imposed displacements, in which case $f_i^{ext}=0$. In the stationary one-dimensional case, the phase-field exhibits a hyperbolic tangent profile. This defines an intrinsic length scale
\[
l_0 = \sqrt{\frac{\kappa_1}{\kappa_2}},
\]
which controls the width of the diffuse interface and thus sets the numerical resolution of the fracture zone.

All simulations are performed using the open-source COMFIT package \cite{skogvoll2024}, based on operator splitting and spectral methods under periodic boundary conditions. Time integration is performed with a fourth-order Runge-Kutta scheme using a uniform dimensionless grid spacing $\Delta x = \Delta y = 1$ and  timestep $\Delta t = 0.01$ which ensures numerical stability.
\subsection{Domain geometry}
\begin{figure}[t!]
  \centering
  \begin{subfigure}[b]{0.49\linewidth}
  \centering
    \includegraphics[width=1\linewidth]{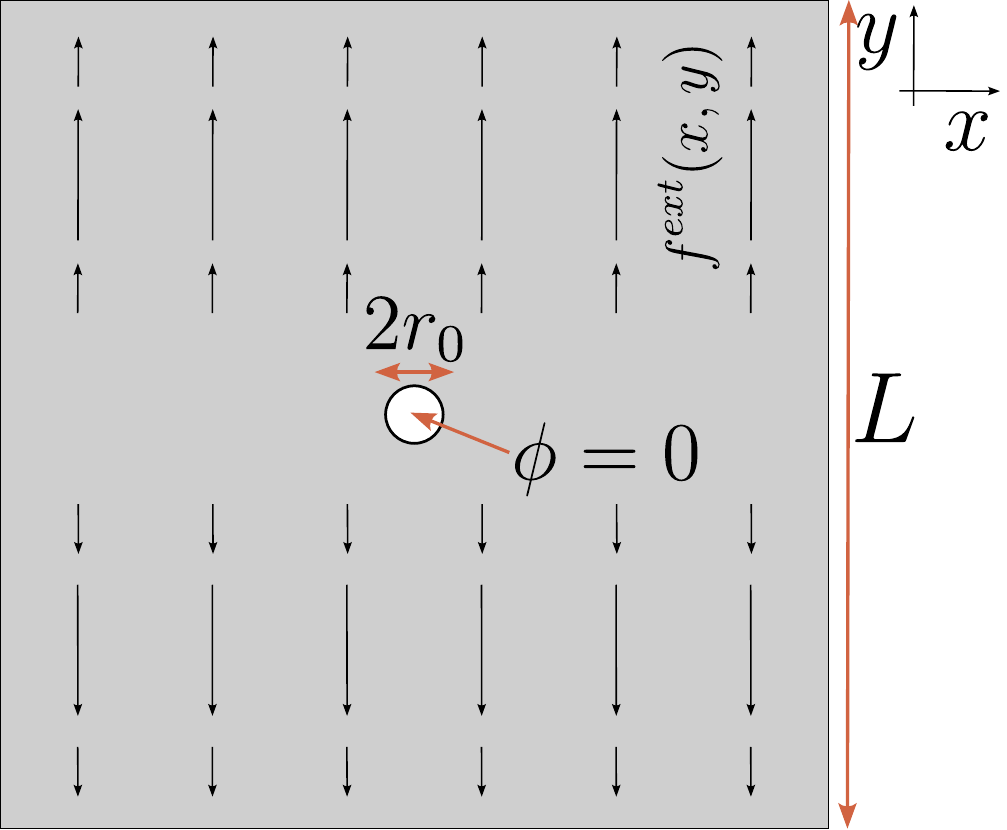}
    \put(-15,0){\textbf{(a)}}
    \phantomcaption
    \label{fig:forcings_a}
  \end{subfigure}
  \hfill
  \begin{subfigure}[b]{0.49\linewidth}
    \centering
    \includegraphics[width=1\linewidth]{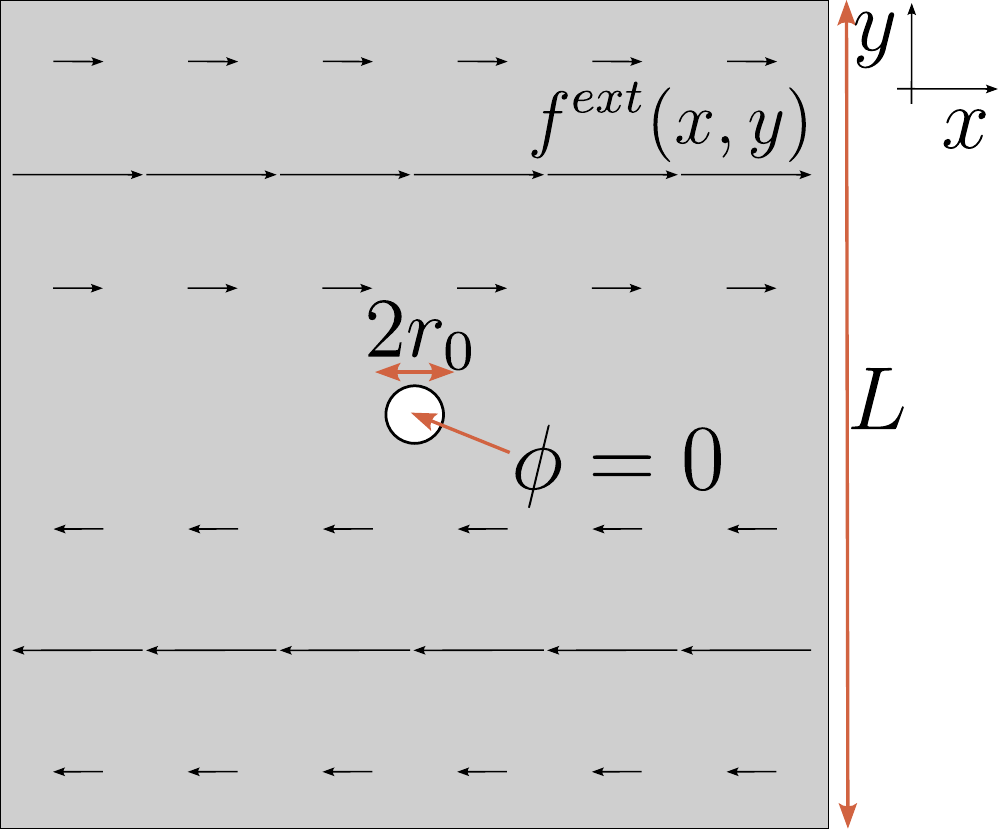}
    \put(-15,0){\textbf{(b)}}
    \phantomcaption
    \label{fig:forcings_b}
  \end{subfigure}
  \caption{Loading and boundary conditions on a two-dimensional elastic domain with a circular inclusion for (a) the tensile and (b) the simple shear numerical experiments. The inclusion has a phase-field value $\phi=0$ and a radius $r_0$. The horizontal dimension of the square domain is $L$. A body force $f^{ext}$ is applied to the system.}
  \label{fig:forcings}
\end{figure}
We consider three basic loading configurations: two cases in which external forcing is applied as a body force, and a third case in which loading is imposed through prescribed boundary displacements.
\subsubsection{Body force loading: uniaxial tension and simple shear experiments}
\label{body_force_experiment_description}
The computational domain is a square region initially filled with intact material, i.e., $\phi=1$, representing a continuous sea-ice cover. The external forcing is taken to vary only along the $y$-direction as:
\begin{equation}
\vec f_{ext}(y) = F_0\, y(y_{max}-y)\, \tanh\!\left(\frac{y - y_{mid}}{y_{max}}\right)\,\vec e,
\label{eq:f_ext}
\end{equation}
where $\vec e$ is the unit vector in the direction of loading, taken either along the $y$-axis for uniaxial tension (Figure~\ref{fig:forcings_a}) or along the $x$-axis for simple shear (Figure~\ref{fig:forcings_b}). The parameter $F_0$ controls the overall forcing amplitude and thus the loading intensity.

This spatial form produces a smooth forcing profile that changes sign at $y=y_{mid}$, corresponding to the mid-height of the domain. It generates a gradient that is strongest in the central region at $y=y_{max}(\frac{1}{2} \pm \frac{1}{4})$. The force vanishes at the boundaries, thereby remaining compatible with periodic boundary conditions required by spectral methods.

The body force represents wind-induced surface stress, which is the dominant driver of sea-ice deformation on short to intermediate timescales (from seconds up to weeks). The system is initially prepared in a fully intact state, $\phi=1$ everywhere. After a short transient during which elastic stresses relax toward a quasi-steady configuration, a circular inclusion is introduced at the center of the domain, where the phase-field is set to $\phi=0$, representing pre-existing fractured or weakened ice, as illustrated in Figure~\ref{fig:forcings}.
\subsubsection{Annular Couette flow}
\label{sec:description_annular_couette_setup}
We consider an annular Couette configuration, in which shear deformation is imposed through a prescribed displacement at the inner boundary rather than via a body force. This setup is of interest for several reasons: (i) the annular geometry produces a deformation that is homogeneous in the azimuthal direction but strongly localized in the radial direction, consistent with the directional localization observed in sea-ice shear bands (see Figure~\ref{fig:description_of_ice}); (ii) it allows the application of large, controlled deformations locally; and (iii) an analytical solution exists for the displacement field in the linear elastic, isotropic limit. In addition, this geometry enables direct comparison with laboratory Couette experiments on floating ice plates, where fracture is induced by rotating an inner cylinder and measuring the resulting torque \cite{cohesion2016,floating_ice_plate}.
\begin{figure}
    \centering
    \includegraphics[width = 0.5\linewidth]{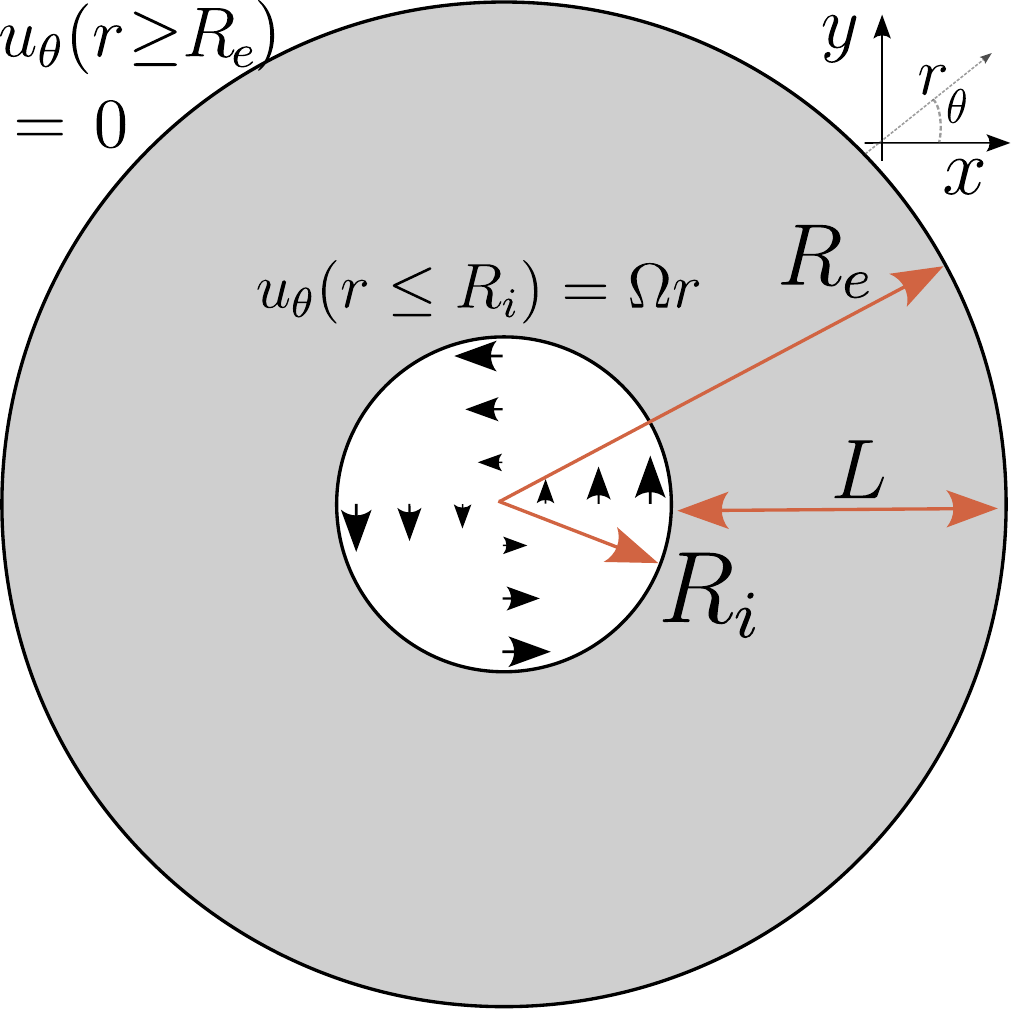}
    \caption{Loading in the Couette configuration. A constant, non-zero azimuthal displacement is prescribed within the inner cylinder, and thus, along the inner boundary of the annulus. The azimuthal displacement is set to zero in the outer cylinder and thus, along the outer boundary of the ice annulus. Initially, the ice cover (in gray) is intact, with $\phi_{t=0} = 1$. After relaxation of the displacement field, it is then free to evolve. The elementary rotation angle is represented by $\Omega$.}
    \label{fig:couette_setup_description}
\end{figure}

The computational domain is illustrated in Figure~\ref{fig:couette_setup_description}. The inner cylinder has radius $R_i$ and is prescribed an azimuthal displacement corresponding to a rotation angle $\Omega$, while the outer cylinder of radius $R_e$ is held fixed. Because the deformation is purely azimuthal and depends only on the radial coordinate $r$, the displacement field is written as $u_\theta(r)$.

For an intact homogeneous material ($\phi=1$), the equilibrium solution of linear elasticity under the boundary conditions $u(R_i)=\Omega R_i$ and $u(R_e)=0$ yields the classical Couette profile \cite{landau1986}:
\begin{equation}
    u(r) = \frac{\Omega R_i^2}{R_i^2 - R_e^2}\left(\frac{r^2 - R_e^2}{r}\right).
    \label{eq:couette_analytical_solution}
\end{equation}
This solution shows that both displacement and strain energy are maximized near the inner boundary, indicating that fracture is expected to initiate preferentially in this region.

In the numerical implementation, the phase-field is initially set to $\phi=1$ throughout the domain and is then allowed to evolve freely. However, within the inner and outer cylinders, $\phi$ is held fixed at $\phi=1$ at all times in order to prevent fracture initiation in these rigid support regions. These zones (shown in white in Figure~\ref{fig:couette_setup_description}) therefore remain intact and serve solely to impose the boundary displacement.

Numerical simulations are performed in three stages. First, the azimuthal displacement of the inner cylinder is gradually increased to a prescribed target value. This value is chosen such that the resulting maximum strain energy density in the intact material near the inner boundary approaches the fracture threshold. A detailed analysis of the resulting stress and strain distributions is presented in Section~\ref{sec:couette_results}. Second, the system is relaxed until the displacement field converges to the analytical Couette solution given in Eq.~\eqref{eq:couette_analytical_solution}. Finally, fracture is activated by coupling this equilibrated displacement field to the phase-field evolution.
\subsection{Dimensionless system of equations}
To reduce the number of independent control parameters, we non-dimensionalize the governing equations. We use as reference scales: i) $L$ the characteristic size of the domain; ii) $c_R$ the Rayleigh elastic wave speed of the intact material; iii) $Y$ the Young's modulus of the intact material; iv) $h$ the average thickness of the thin ice plate. Then, we rescale space, time, operators, external body force, internal stress and strain energy in a dimensionless form as in Table \ref{tab:nondim_vars}:
\begin{table}[h!]
  \raggedright
  \renewcommand{\arraystretch}{1.5}
  \begin{tabularx}{0.48\textwidth}{
    l
    >{\raggedright\arraybackslash}m{1.8cm}
    l
}
    \toprule
    Quantity                    & \shortstack[l]{Dimensional\\symbol}
                                & Scaling \\
    \midrule
    Time                        & $t$  
                                & $\tilde{t} = t \frac{c_R}{L}$ \\
    Position                    & $x$ 
                                & $\tilde{x} = \frac{x}{L}$ \\
    Time derivative             & $\partial_{t}$ 
                                & $\tilde{\partial_{t}}= \frac{L}{c_R}\,\partial_{t}$ \\
    Gradient                    & $\nabla$ 
                                & $\tilde{\nabla} = L\nabla$ \\
    Elastic strain energy       & $\mathscr{E}_{\mathrm{strain}}$ 
                                & $\tilde{\mathscr{E}}_{\mathrm{strain}} 
                                   = \mathscr{E}_{\mathrm{strain}}\frac{(1+\nu)}{Yh}$ \\
    Internal stress             & $\sigma$  
                                & $\tilde{\sigma} = \frac{\sigma}{Y}$ \\
    External body force         & $f_{\mathrm{ext}}$ 
                                & $\tilde{f}_{\mathrm{ext}} 
                                   = f_{\mathrm{ext}}\frac{ L}{Y}$ \\
    Displacement field          & $u$ 
                                & $\tilde{u} = \frac{u}{L}$ \\
    Interface thickness         & $l_{0}$ 
                                & $\tilde{l}_{0} = \frac{l_{0}}{L}$ \\
    \bottomrule
  \end{tabularx}
    \caption{Dimensional and dimensionless variables and their scaling relationships.}
  \label{tab:nondim_vars}
\end{table}
Inserting the rescaled variables into Eq.~\ref{eq:evolution_equation_PF} yields the dimensionless form of the phase-field evolution equation:
\begin{equation}
    \begin{aligned}
    \tilde{\partial_t}\phi &= \left(\tilde{l_0}^2\frac{\chi L \kappa_2}{c_R}\tilde{\nabla}^2\phi -  \frac{\chi L \kappa_2}{c_R}V'_{DW}(\phi) - \frac{\chi L}{c_R} \frac{Yh}{1+\nu}g'(\phi) \Delta \mathscr{E} \right), 
    \end{aligned}
\end{equation}
where $\Delta \mathscr{E} = \tilde{\mathscr{E}}_{\text{strain}}-\frac{\mathscr{E}_c(1+\nu)}{Yh}$.\\
For the momentum balance equation, rescaling of stress, force and displacement are applied, which transforms the overdamped momentum balance into:
\begin{equation}
 \tilde{\partial}_t \tilde{u}_i = \frac{Y}{\beta \rho L c_R} \left ( \tilde{\partial}_j \tilde{\sigma}_{ij} + \tilde{f_i}^{ext}\right).
 \label{eq:displacement_evolution_dimensionless}
\end{equation}
To simplify the notations, in the following, we omit the tilde on dimensionless variables. The pre-factors form a set of five dimensionless parameters $\{N_1, N_2, N_3, N_4, N_5\}$ defined in Table \ref{tab:dimensionless_parameters}. Using these parameters, the dimensionless governing equations write:
\begin{equation}
\begin{cases}
    \partial_t\phi &= \left[N_1\nabla^2\phi -  N_2V'_{DW}(\phi) - N_3g'(\phi) \left(\mathscr{E}_{\text{strain}}-N_4\right)   \right] \\
    \partial_tu_i &= N_5 \left ( \partial_j\sigma_{ij} + f^{ext}_i\right)
\end{cases}.
\label{eq:evolution_equation_dimensionless}
\end{equation}
The ratio $\frac{N_1}{N_2}$ is equal to $l_0^2$ and is used to calibrate the thickness of the interface. The timescale separation is governed by the parameter $N_5$. Indeed a relatively small value of $\beta$ is chosen so that mechanical relaxation is represented as rapid, although further reduction is limited by the need for increasingly small time-step resolution.
\begin{table}[h]
\centering
\renewcommand{\arraystretch}{1.8}
\setlength{\tabcolsep}{2pt}
\begin{tabularx}{0.5\textwidth}{l l X}
\hline
Parameter & $\,$ Formula & $\,$ Controls \\
\hline
$N_1$ & $l_0^2\dfrac{\chi L \kappa_2}{c_R}$ 
& Interfacial width \\
$N_2$ & $\dfrac{\chi L \kappa_2}{c_R}$ 
& Energy barrier for $\phi$ \\
$N_3$ & $\dfrac{\chi L}{c_R}\dfrac{Yh}{1+\nu}$ 
& Elastic energy \\
$N_4$ & $\dfrac{(1+\nu)\mathcal{E}_c}{Yh}$ 
& Critical energy \\
$N_5$ & $\dfrac{Y}{\beta \rho L c_R}$ 
& Displacement relaxation \\
%\hline
$\nu$ & $\nu$
& Poisson's ratio \\
\hline
\end{tabularx}
\caption{Dimensionless model parameters and their physical interpretation.}
\label{tab:dimensionless_parameters}
\end{table}
\section{Results}
\subsection{Tensile loading}\label{sec3}
\subsubsection{Stress profile before fracture}
We first validate the model by comparing the computed stress field with the classical analytical solution of the Kirsch's problem \cite{kirsch1898}. Some deviations are expected, since our numerical setup differs from the idealized assumptions of infinite geometry and uniform far-field loading.

The Kirsch's problem is a benchmark solution in linear elasticity that describes the stress distribution in an infinite, homogeneous, isotropic elastic plate containing a traction-free circular hole. The system is subjected to a uniform uniaxial tensile stress $\sigma_0$ applied at infinity (here taken in the $y$-direction). The solution captures how the presence of the hole distorts the stress field and generates a strong stress concentration at the hole boundary, where the maximum stress can reach up to three times the applied far-field load.

\begin{figure*}[!t]
    \centering
    \begin{subfigure}{\linewidth}
        \centering
        \includegraphics[width=0.9\linewidth]{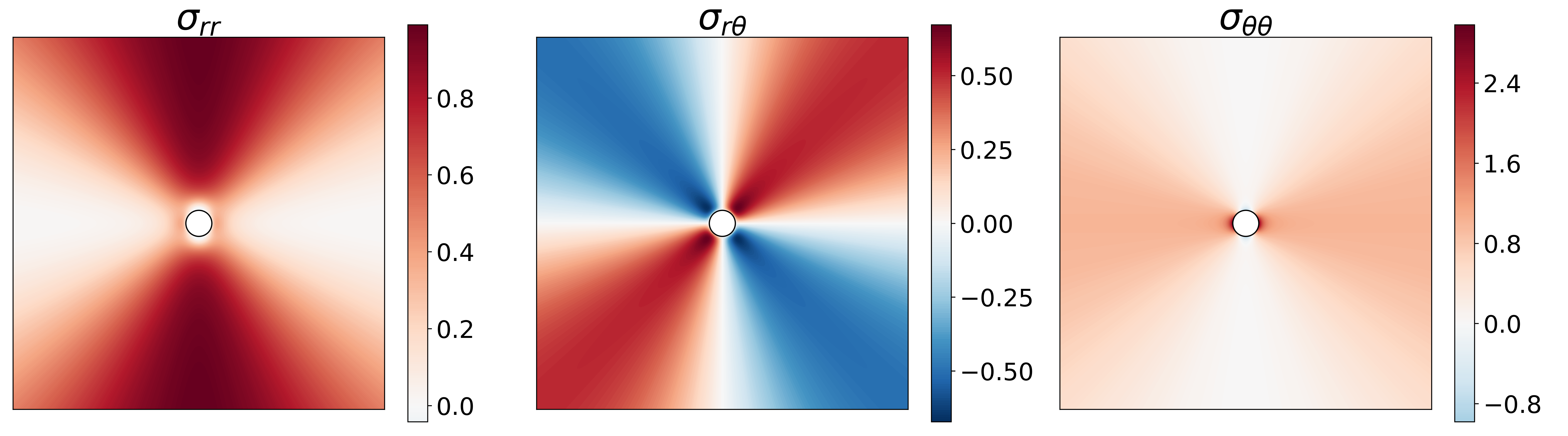}
        % \caption{}
        \put(-480,60){\textbf{(a)}}
        \phantomcaption
        \label{fig:stress_around_incl_a}
    \end{subfigure}
    % \vspace{5cm} % Optional space between figures
    \begin{subfigure}{\linewidth}
        \centering
        \includegraphics[width=0.9\linewidth]{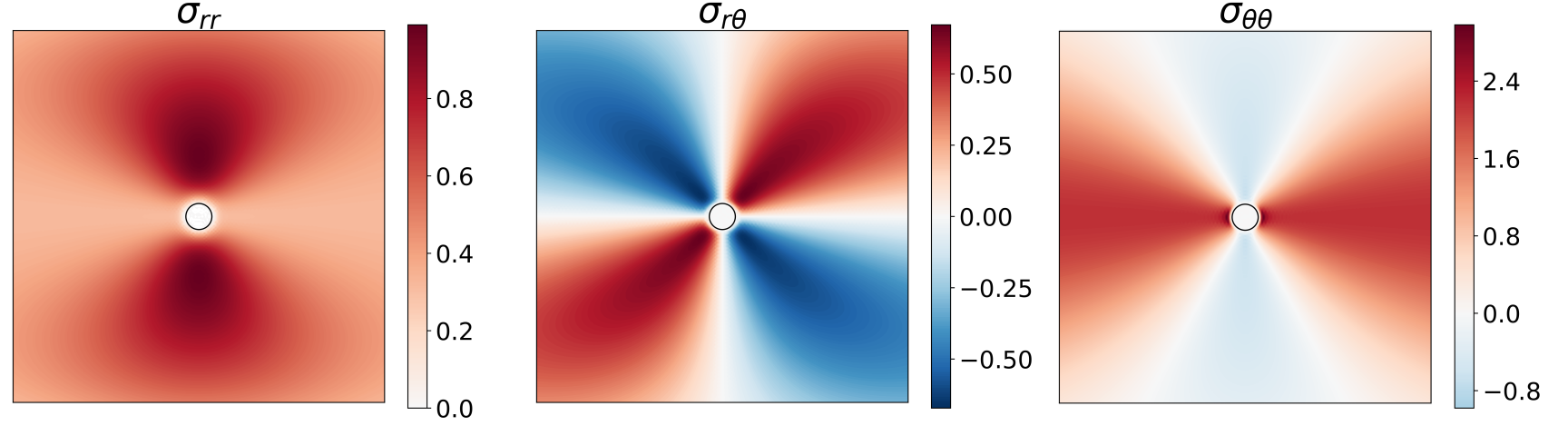}
        % \caption{}
        \put(-480,60){\textbf{(b)}}
        \phantomcaption
        \label{fig:stress_around_incl_b}
    \end{subfigure}
    \caption{(a) Kirsch's analytical solution of the stress field around a circular inclusion, in polar coordinates, and (b) corresponding numerical simulation of the stress field for a linear elastic matrix surrounding a circular inclusion. This central square cross-section has dimensions $200\times 200$ elements and is part of a larger simulation domain with dimensions $500\times200$ elements. To mitigate Gibbs oscillations, the edges of the central inclusion have been smoothed using a Gaussian filter. The simulation considers only the elastic response by activating the displacement field evolution equation, while the phase-field evolution equation remains uncoupled.}
    \label{fig:stress_around_incl}
\end{figure*}

Within linear elasticity, the resulting equilibrium stress field around the hole is obtained in polar coordinates $(r,\theta)$ as:
\begin{equation}
    \begin{aligned}
    \sigma_{rr} &=\frac{\sigma_0}{2}\left(1- \frac{a^2}{r^2} \right)-\frac{\sigma_0}{2} \left(1+3\frac{a^4}{r^4}-4\frac{a^2}{r^2} \right) \cos(2\theta) \\
    \sigma_{\theta \theta} &= \frac{\sigma_0}{2} \left(1+\frac{a^2}{r^2} \right)+\frac{\sigma_0}{2}\left(1+3\frac{a^4}{r^4} \right)\cos(2\theta)\\
    \sigma_{r\theta} &= \frac{\sigma_0}{2}\left(1-3\frac{a^4}{r^4}+2\frac{a^2}{r^2} \right)\sin(2\theta)
    \end{aligned},
    \label{eq:kirsch_solution_stress}
\end{equation}
where $r$ is the radial distance from the center of the hole and $\theta$ is the polar angle measured from the $x$-axis. The angular profiles of the individual stress components $\sigma_{xx}, \sigma_{yy}, \sigma_{xy}$ are displayed in Figure~\ref{fig:stress_around_incl_a}.
We reproduce the Kirsch's configuration within our numerical framework by applying a body force that generates an effective uniaxial tensile loading (Figure~\ref{fig:forcings_a}). However, two key differences from the classical analytical problem must be noted. First, the circular inclusion is not strictly traction-free, since the phase-field formulation introduces a finite interfacial transition zone. Second, periodic boundary conditions lead to finite-size effects, particularly near the top and bottom boundaries where the imposed body force vanishes.

To reduce these artifacts, the computational domain is elongated in the loading direction, and the comparison with the analytical solution is restricted to a central region around the inclusion. Within this region, the computed stress field (Figure~\ref{fig:stress_around_incl_b}) reproduces the characteristic angular structure of the Kirsch's solution, including its expected symmetries. This confirms that the elastic solver correctly captures the stress concentration induced by a circular inclusion. Small deviations in the radial profile can be attributed to the spatial structure of the imposed body force and the influence of periodic boundaries.
\subsubsection{Fracture propagation}
\begin{figure}[!t]
  \centering
  \begin{subfigure}[t]{0.63\textwidth}
      \includegraphics[width=0.8\linewidth]{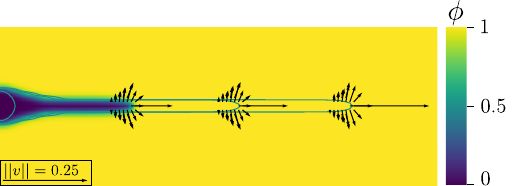}
      \phantomcaption
      \put(-280,0){\textbf{(a)}}
      \label{fig:tensile_propagation_isolines_a}
  \end{subfigure}
  \hfill
  \begin{subfigure}[t]{1\textwidth}
      \includegraphics[width=0.49\linewidth]{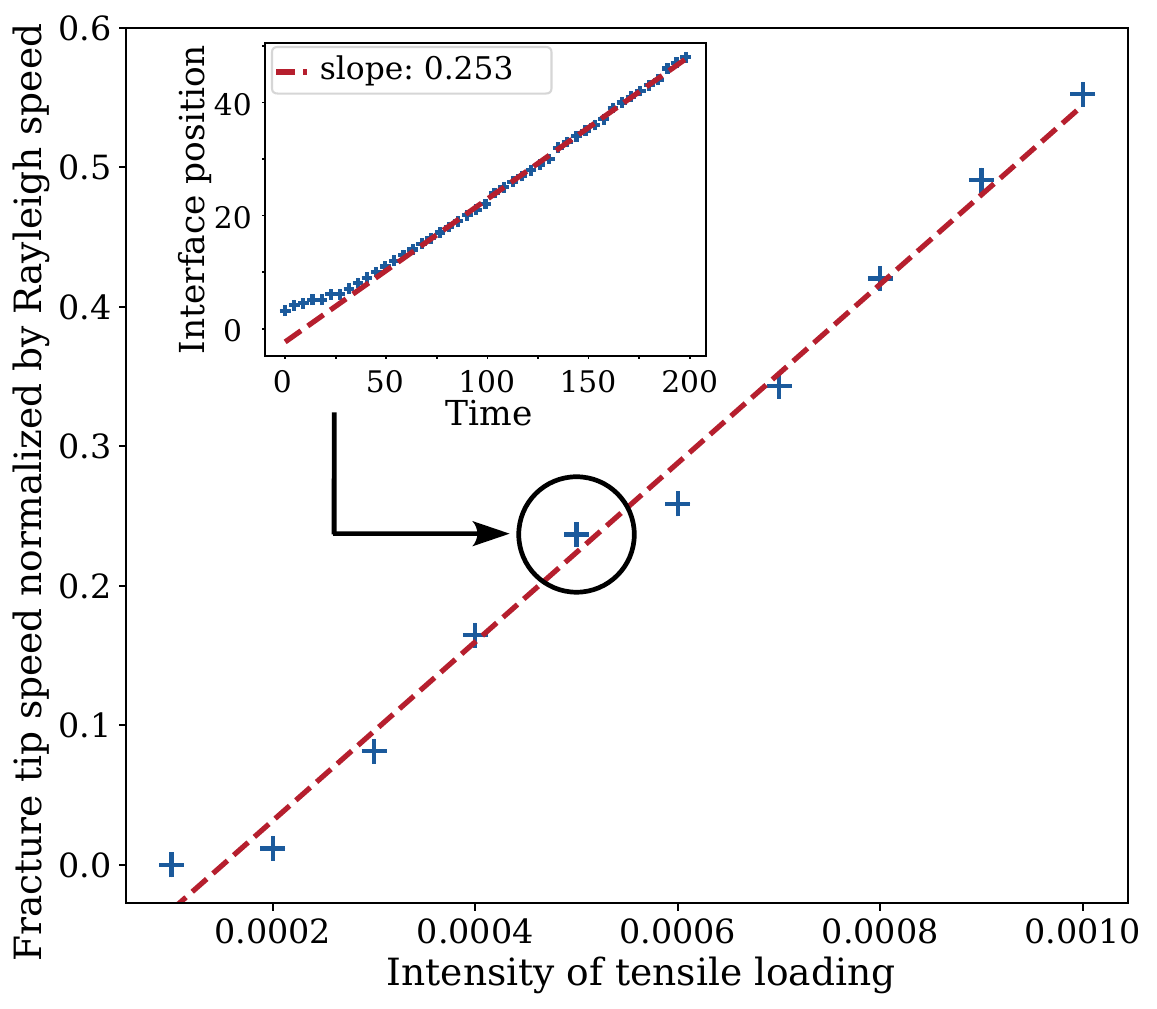}
      \phantomcaption
      \put(-265,15){\textbf{(b)}}
      \label{fig:tensile_propagation_isolines_b}
  \end{subfigure}
  \caption{Mode-I fracture under tensile loading. (a) Propagation of a mode-I fracture nucleating from an inclusion under tensile loading. The blue lines are the isolines of $\phi =0.5$ at different time steps. The velocity field is represented by arrows at each time step. (b) Crack propagation speed as function of the intensity of external loading. Simulations are represented with pluses and the dashed-red line show a linear fit. Inset shows one simulation of the position of the fracture tip as a function of time and the slope (red line) is the calculated velocity.}
  \label{fig:tensile_propagation_isolines}
\end{figure}
We now use the stress field obtained in the previous section as the initial condition to study mode-I fracture propagation within the phase-field framework. In classical fracture mechanics, propagation is governed by Griffith’s criterion, which states that a crack advances when the elastic energy release rate exceeds the fracture energy \cite{griffith1921}. In the present formulation, this condition is naturally incorporated through the coupling between the phase-field and the local strain energy density, so that fracture develops when the elastic energy locally surpasses the critical threshold.

Accordingly, we solve the coupled system defined in Eq.~\ref{eq:evolution_equation_dimensionless} and analyze the resulting crack dynamics. Two aspects are of particular interest: (i) whether the crack propagation direction is consistent with the applied loading, and (ii) whether the model reproduces Griffith-like behavior, in which the crack speed increases with the applied load near the propagation threshold.

Under uniaxial tension, fracture is expected to propagate perpendicular to the loading direction. This follows from linear elastic fracture mechanics, where mode-I cracks align normal to the maximum principal tensile stress \cite{anderson2005}. To illustrate this behavior, we exploit symmetry and present results for one half of the domain in Figure~\ref{fig:tensile_propagation_isolines_a}. The crack is identified using the phase-field contour $\phi = \phi_c = 0.5$, shown at successive times.

To quantify crack propagation, we track the crack tip as the point on the $\phi=0.5$ contour that is farthest from the initial inclusion center, and compute its velocity via numerical differentiation in time. The normal velocity is determined by the phase-field evolution as $v_n = -\frac{\partial_t \phi}{\|\nabla \phi\|}$, as detailed in, e.g., Ref.~\cite{skogvollUnifiedFieldTheory2023}. This expression is used to reconstruct the crack speed field at different stages of propagation, as shown in Figure~\ref{fig:tensile_propagation_isolines_a}.

The crack speed is calculated for different values of loading amplitude $F_0$ near the critical load required for crack propagation, as shown in Figure \ref{fig:tensile_propagation_isolines_b}. For subcritical loading ($F_0<2.10^{-4}$), the strain energy remains below the fracture threshold, and no fracture propagates. For overcritical loading, as loading increases ($3.10^{-4}<F_0<1.10^{-3}$), the crack speed increases linearly with the applied force. This linear regime, obtained for loading conditions close to the critical load for fracture propagation, is expected, as further justified by the analytical derivation below, and provides a benchmark for the behavior of the numerical model. In the simulations, the crack speed is computed in units of the Rayleigh speed and $v_n<1$, consistent with the expected sub-Rayleigh propagation regime.

The linear dependence of crack speed on external loading can be derived analytically. For a steadily propagating crack with normal velocity $v_n$, we rewrite the phase-field evolution equation (Eq.~\ref{eq:evolution_equation_PF}) in a co-moving frame attached to the interface. Introducing the coordinate $\xi$ normal to the interface, we obtain
\[
\partial_t = -v_n \partial_\xi, \qquad \nabla = \partial_\xi, \qquad \nabla^2 = \partial_\xi^2,
\]
so that the steady-state equation in co-moving frame becomes
\begin{equation}
-v_n \phi'(\xi)
= \chi\left(\kappa_1 \phi''(\xi) - \kappa_2 V'_{DW}(\phi) - g'(\phi)\Delta E\right),
\end{equation}
with $\phi' = d\phi/d\xi$. To obtain the crack speed, we multiply by $\phi'$ and integrating across the interface. The gradient contribution vanishes because $\phi' \to 0$ as $\xi \to \pm\infty$:
\begin{equation}
\int_{-\infty}^{+\infty} \kappa_1 \phi'' \phi' \, d\xi = 0.
\end{equation}
The same holds for the double-well term, since the interface connects two equal-energy minima:
\begin{equation}
\int_{-\infty}^{+\infty} \kappa_2 V'_{DW}(\phi)\phi' \, d\xi = 0.
\end{equation}
Thus, only the coupling to the elastic energy remains:
\begin{equation}
\int_{-\infty}^{+\infty} \Delta E\, g'(\phi)\phi' \, d\xi
= \Delta E \left[g(1)-g(0)\right] = \Delta E.
\end{equation}
This yields the steady-state relation:
\begin{equation}
v_n = \chi \frac{\Delta E}{\int_{-\infty}^{+\infty} (\phi')^2 d\xi}.
\label{eq:relation_speed_deltaE}
\end{equation}
The denominator is a positive finite constant set by the intrinsic structure of the diffuse interface, which connects two distinct minima over a finite width. Equation~\eqref{eq:relation_speed_deltaE} shows that the crack speed is directly proportional to the excess driving energy $\Delta E$. In the simulations, this excess is controlled by the applied load, leading to the behavior observed in Figure~\ref{fig:tensile_propagation_isolines_b}, where the crack speed increases linearly near the propagation threshold. Although the elastic energy scales quadratically with loading in linear elasticity, the relevant driving quantity is the energy above threshold. Close to the critical load $F_c$, it can be approximated as
\[
\Delta E \sim F_0^2 - F_c^2 \approx 2F_c(F_0 - F_c),
\]
which is linear in the distance from threshold. This scaling directly explains the observed linear dependence of crack speed on loading in the near-critical regime.
\subsection{Simple, plane shear loading}
\begin{figure*}[tbp]
    \centering
    \begin{subfigure}[t]{0.21\textwidth}
        \centering
        \includegraphics[width=\linewidth]{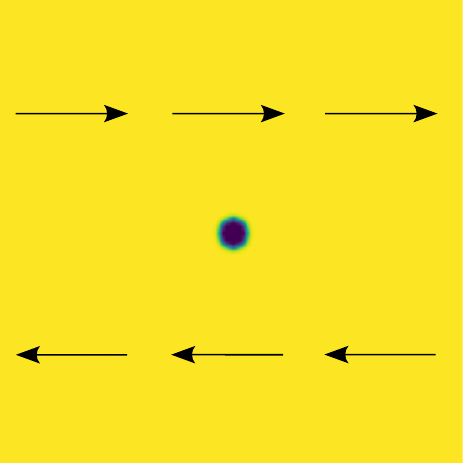}
        \phantomcaption
        \put(-130,50){\textbf{(a)}}
        \label{fig:propagation_simple_shear_a}
    \end{subfigure}
    \hfill
    \begin{subfigure}[t]{0.21\textwidth}
        \centering
        \includegraphics[width=\linewidth]{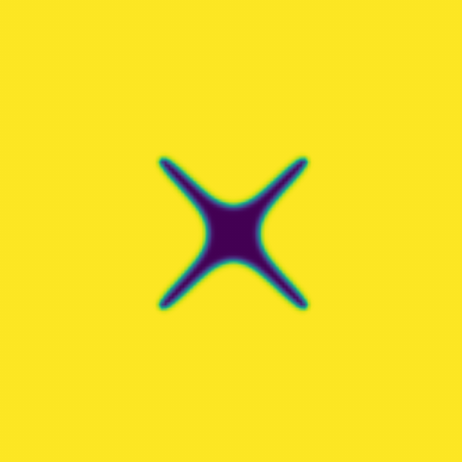}
    \end{subfigure}
    \hfill
    % \vspace{0.5cm}
    \begin{subfigure}[t]{0.21\textwidth}
        \centering
        \includegraphics[width=\linewidth]{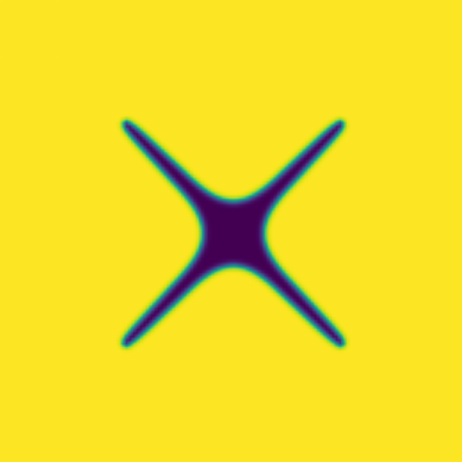}
    \end{subfigure}
    \hfill
    \begin{subfigure}[t]{0.21\textwidth}
        \centering
        \includegraphics[width=\linewidth]{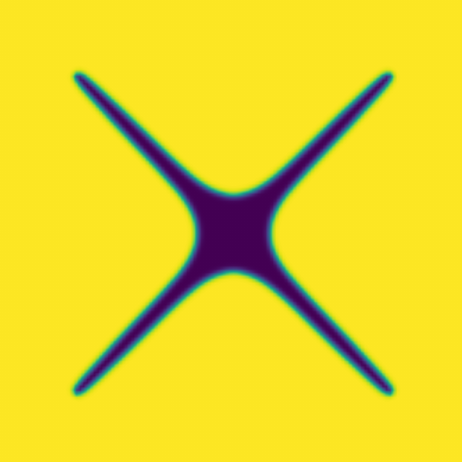}
    \end{subfigure}
    \hfill
    \begin{subfigure}[t]{0.06\textwidth}
        \centering
        \includegraphics[width=\linewidth]{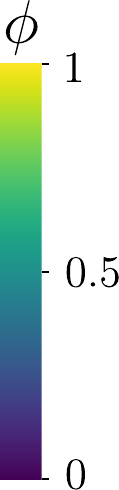}
    \end{subfigure}
    \\[0.3cm]
    \begin{subfigure}[t]{0.21\textwidth}
        \centering
        \includegraphics[width=\linewidth]{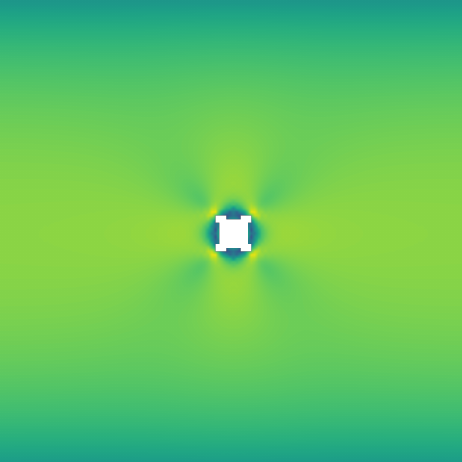}
        \phantomcaption
        \put(-130,50){\textbf{(b)}}
        \label{fig:propagation_simple_shear_b}
    \end{subfigure}
    \hfill
    \begin{subfigure}[t]{0.21\textwidth}
        \centering
        \includegraphics[width=\linewidth]{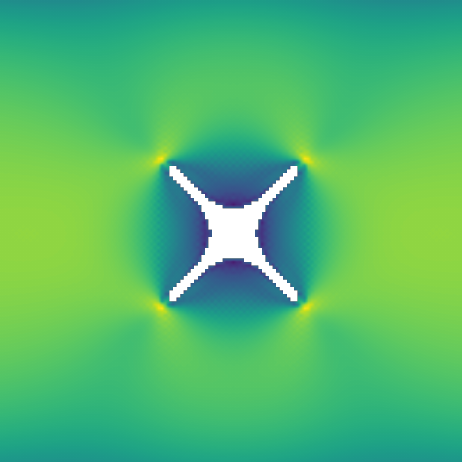}
    \end{subfigure}
    \hfill
    % \vspace{0.5cm}
    \begin{subfigure}[t]{0.21\textwidth}
        \centering
        \includegraphics[width=\linewidth]{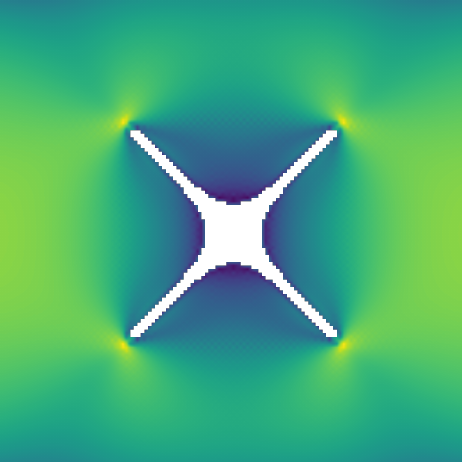}
    \end{subfigure}
    \hfill
    \begin{subfigure}[t]{0.21\textwidth}
        \centering
        \includegraphics[width=\linewidth]{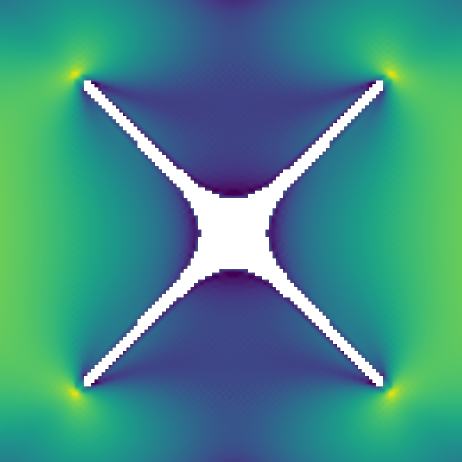}
    \end{subfigure}
    \hfill
    \begin{subfigure}[t]{0.06\textwidth}
        \centering
        \includegraphics[width=\linewidth]{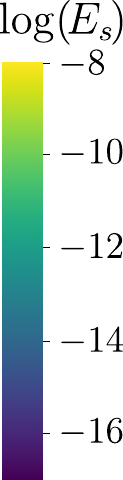}
    \end{subfigure}
    \caption{(a) Evolution of the phase-field during the propagation of four fractures from an inclusion under simple shear loading. The forcing is shown on Fig. \ref{fig:forcings_b}. (b) Corresponding strain energy multiplied by the coupling function of the phase-field, in log scale. The locations where the phase-field is zero are masked.}
    \label{fig:propagation_simple_shear}
\end{figure*}
We now study the propagation of fractures under simple shear loading, with the simulation conditions as described in Section~\ref{body_force_experiment_description} and illustrated in Figure~\ref{fig:forcings_b}. As for the uniaxial tensile loading case, our analysis focuses on  the direction of the fracture propagation and the relation between the crack speed and the magnitude of applied loading.

Figure~\ref{fig:propagation_simple_shear_a} shows the resulting fracture pattern. Four symmetric branches emanate from the central inclusion and propagate along directions approximately aligned with $\pm \pi/4$ and $\pm 3\pi/4$ relative to the shear direction. This orientation is consistent with the principal stress directions in the small-deformation regime. Two of these directions correspond to tensile principal stresses, while the other two correspond to compressive ones.

The associated strain energy field, shown in Figure~\ref{fig:propagation_simple_shear_b}, is plotted in logarithmic scale. At each time step, the strain energy localizes at the crack tips, confirming that propagation is driven by stress concentration at the advancing fronts. It is important to notice that all four fracture branches propagate symmetrically under compression and tension. This symmetry follows directly from the modeling assumptions that the strain energy is treated as a scalar quantity without any tension–compression decomposition, and the fracture criterion depends only on this scalar energy exceeding a threshold. As a result, the model does not distinguish between tensile and compressive failure modes.

In contrast, many geomaterials exhibit strong tension–compression asymmetry, with significantly lower fracture resistance in tension. This behavior can be incorporated through energy decompositions that separate tensile and compressive contributions \cite{henry_levine2004, amor2009, miehe2010}, allowing fracture to be driven primarily by tensile strains. Such extensions are not included here and are left for future work.

\begin{figure*}[!t]
    \centering
    \begin{subfigure}{0.4\linewidth}
        \centering
        \includegraphics[width=\linewidth]{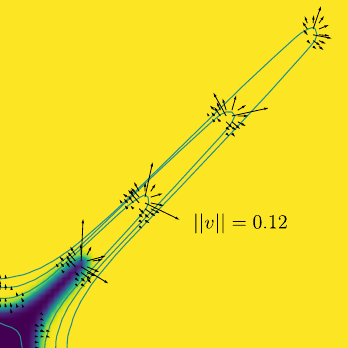}
        \put(0,0){\textbf{(a)}}
        \phantomcaption
        \label{fig:velocity_vector_field_simple_shear_one_branch_a}
    \end{subfigure}
    \hfill
    \begin{subfigure}{0.48\linewidth}
        \centering
        \includegraphics[width=\linewidth]{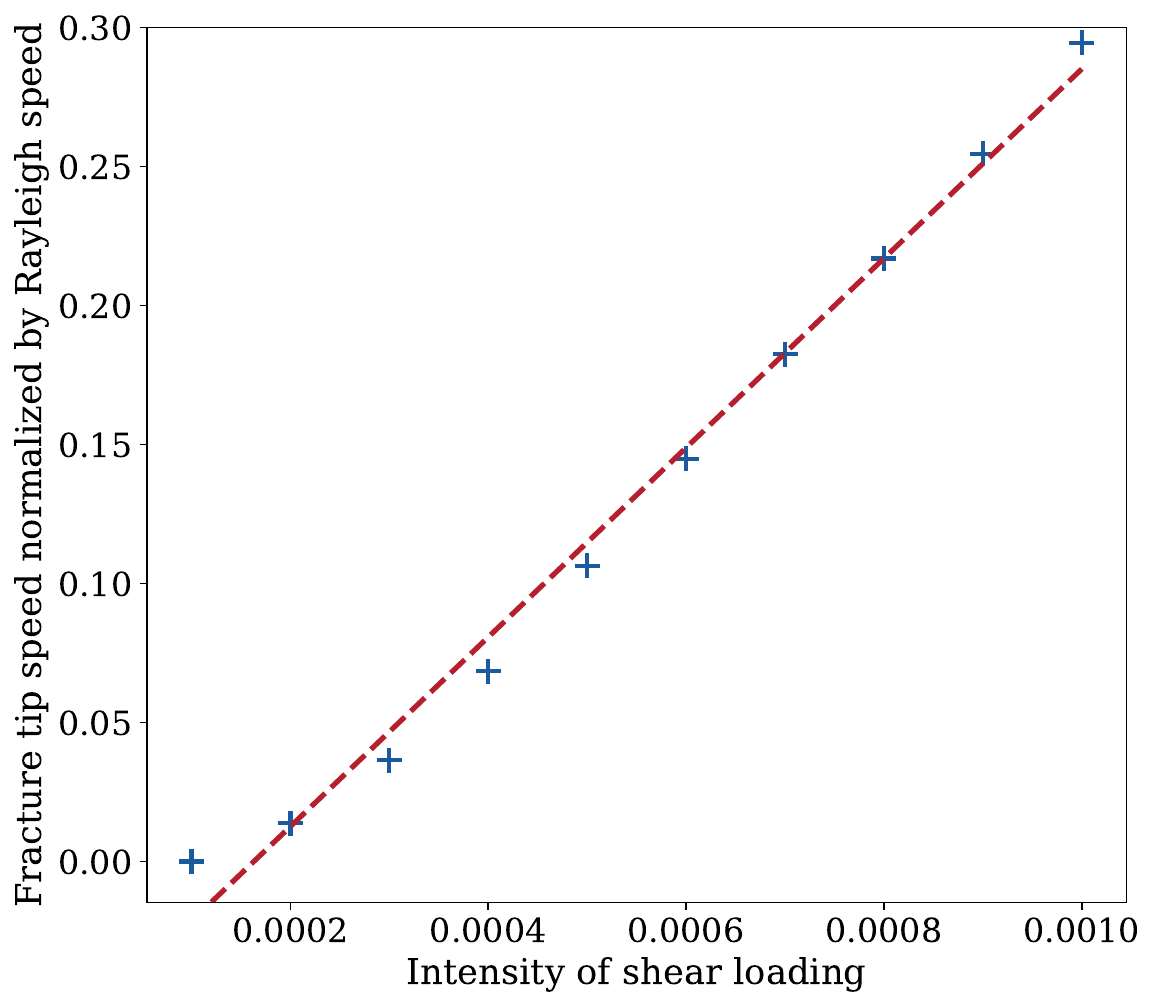}
        \put(0,0){\textbf{(b)}}
        \phantomcaption
        \label{fig:velocity_vector_field_simple_shear_one_branch_b}
    \end{subfigure}
   \caption{(a) Fracture propagation under simple shear, isolines, and velocity vector field. (b) Velocity of the crack tip as a function of the intensity of loading for different simulations (pluses). The red dashed line is a linear fit to the data.} \label{fig:velocity_vector_field_simple_shear_one_branch}
\end{figure*}

Figure~\ref{fig:velocity_vector_field_simple_shear_one_branch_a} shows the velocity field along a representative crack branch, where the vectors indicate both propagation direction and local speed. Small, time-varying deviations from the ideal principal stress orientation are observed due to grid resolution and numerical discretization. The propagation speed is quantified as a function of applied loading magnitude (Figure~\ref{fig:velocity_vector_field_simple_shear_one_branch_b}). As in the tensile case, the crack speed scales linearly with the magnitude of the applied force. However, for comparable fracture energy parameters, crack speeds under simple shear are approximately half those observed under uniaxial tension. This reduction is consistent with the redistribution of available elastic energy across four propagating branches instead of two, effectively reducing the energy flux into each individual crack tip.
\subsection{Simple shear loading in a Couette geometry} \label{sec:couette_results}
Finally, we consider the Couette geometry introduced in Section~\ref{sec:description_annular_couette_setup}. We first compare the pre-fracture elastic displacement field with the analytical solution of linear elasticity, and then analyze the resulting fracture patterns in relation to the simple shear case discussed previously.

\begin{figure}[!t]
    \centering
    \begin{subfigure}[t]{0.49\textwidth}
        \centering
        \includegraphics[width=1\linewidth]{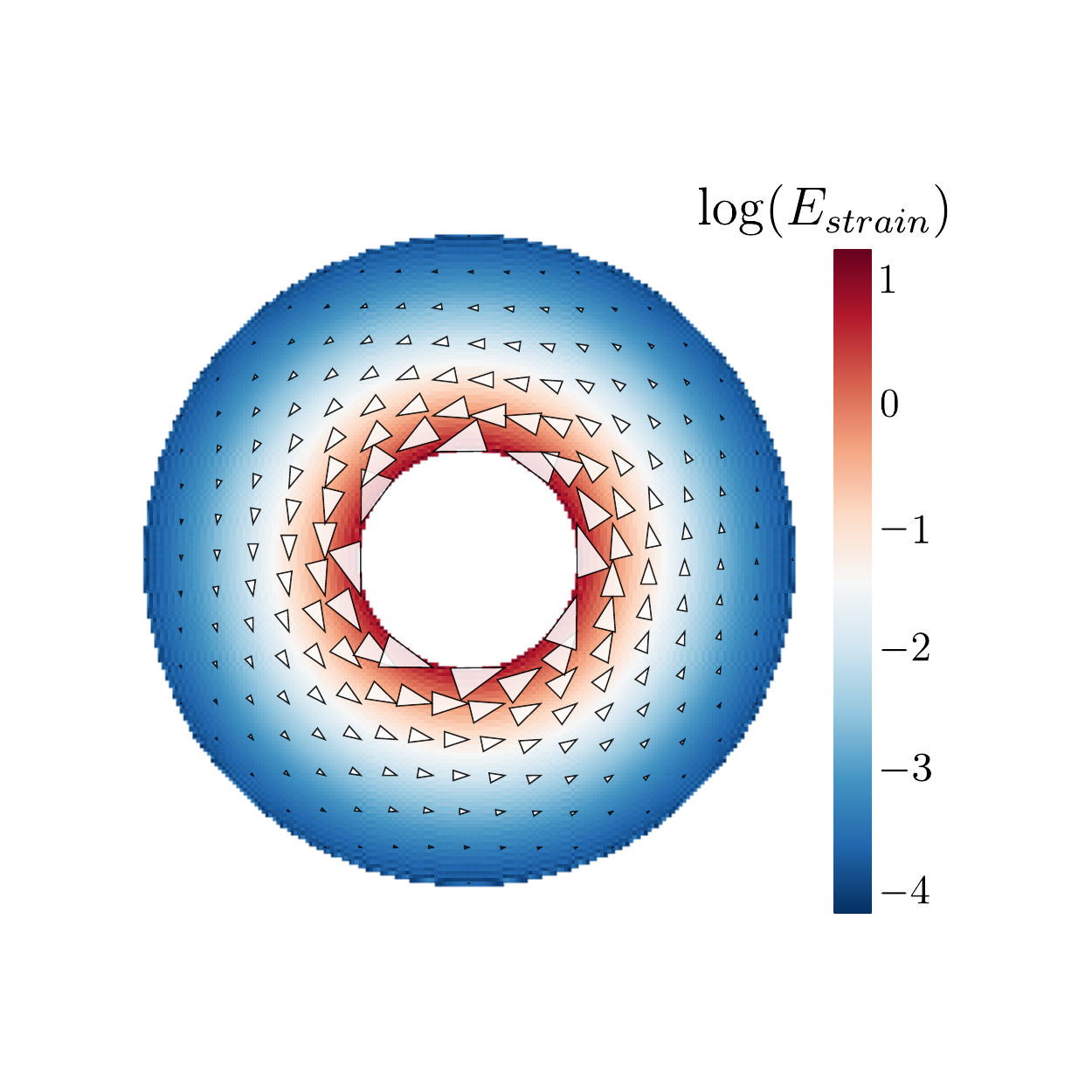}
        \put(-250,120){\textbf{(a)}}
        \phantomcaption
        \label{fig:displacement_and_energy_couette}
    \end{subfigure}
    \hfill
    \begin{subfigure}[t]{0.45\textwidth}
        \centering
        \includegraphics[width=0.8\linewidth]{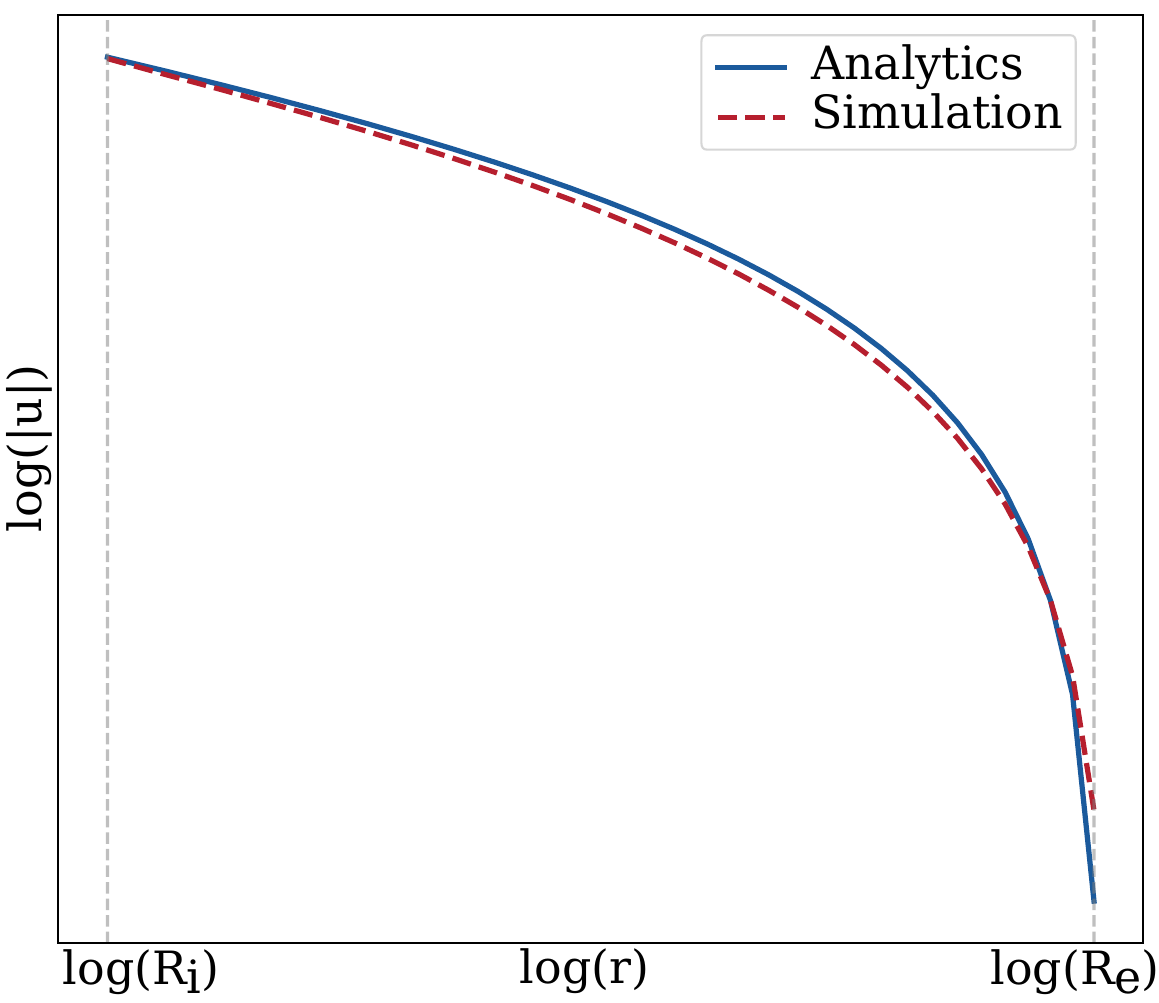}
        \put(-215,80){\textbf{(b)}}
        \phantomcaption
        \label{fig:convergence_displacement_cylinder}
    \end{subfigure}
    \caption{(a) Vector field in the Couette geometry representing the displacement after relaxation, with its amplitude given by the size of the arrows. The color background corresponds to the log of the strain energy before fracturing. (b) Displacement as function of the radius in the ring between the inner and outer cylinders, in log-log scale. The blue line corresponds to the analytical solution (Eq.~ \ref{eq:couette_analytical_solution}) and the dashed, red line corresponds to the simulated displacement after relaxation.}
    \label{fig:before_nucleation_cylinder}
\end{figure}

An azimuthal displacement is first imposed on the inner cylinder, while the outer boundary is fixed. The system is then allowed to relax, during which the displacement field converges radially toward the outer boundary, as shown in Figure~\ref{fig:displacement_and_energy_couette}. After this transient phase, the numerical solution agrees with the analytical Couette profile given in Eq.~\ref{eq:couette_analytical_solution}, as illustrated in Figure~\ref{fig:convergence_displacement_cylinder}.

\begin{figure*}[!t]
    \centering
    \begin{subfigure}[t]{0.2\textwidth}
        \centering
        \includegraphics[width=1\linewidth]{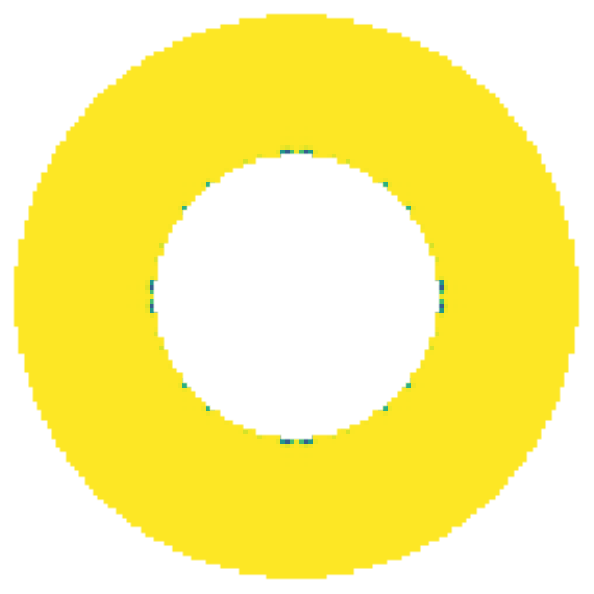}
    \end{subfigure}
    \hfill
    \begin{subfigure}[t]{0.2\textwidth}
        \centering
        \includegraphics[width=1\linewidth]{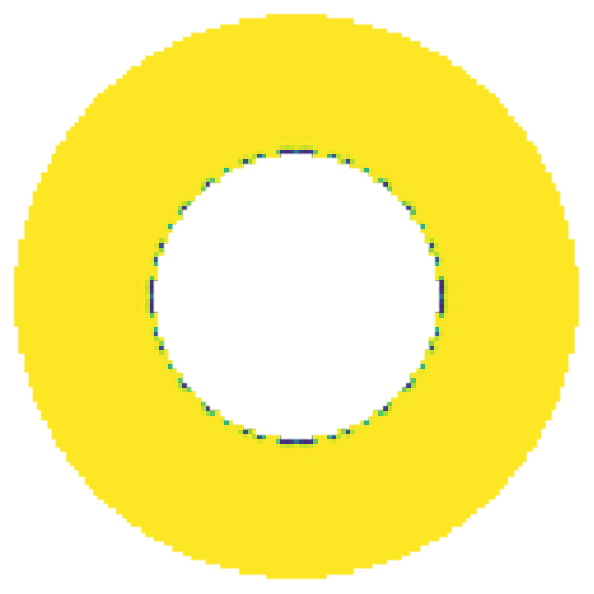}
    \end{subfigure}
    \hfill
    % \vspace{0.5cm}
    \begin{subfigure}[t]{0.2\textwidth}
        \centering
        \includegraphics[width=1\linewidth]{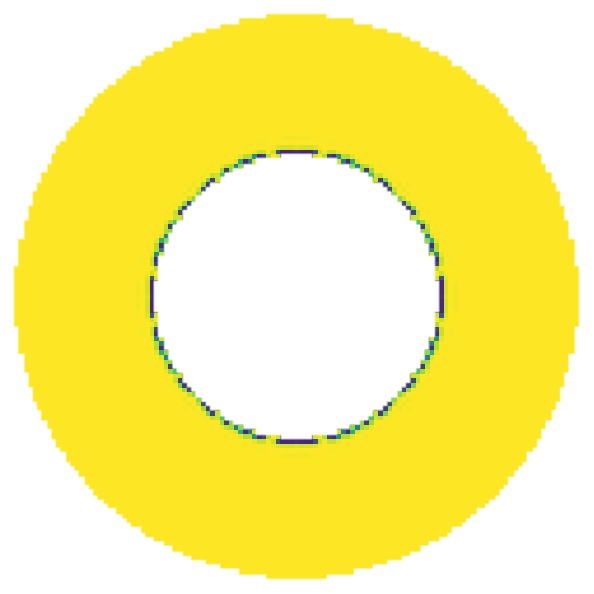}
    \end{subfigure}
    \hfill
    \begin{subfigure}[t]{0.2\textwidth}
        \centering
        \includegraphics[width=1\linewidth]{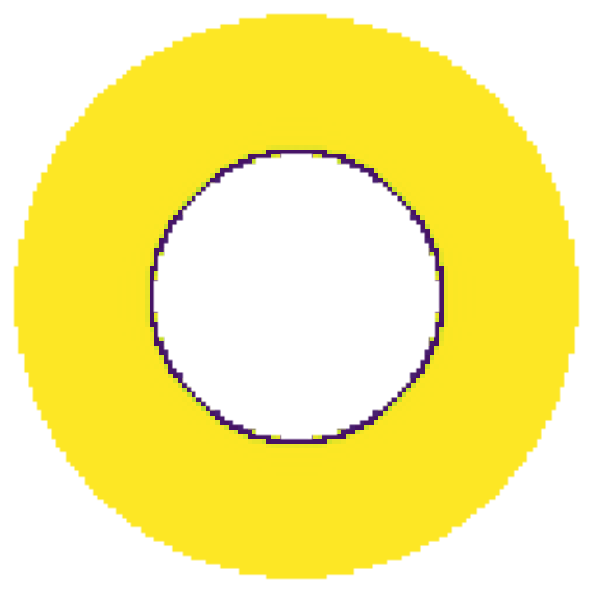}
    \end{subfigure}
        \hfill
    \begin{subfigure}[t]{0.15\textwidth}
        \centering
        \includegraphics[width=0.4\linewidth]{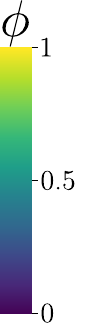}
    \end{subfigure}
    \caption{Fracturing under simple shear loading in the Couette geometry. No front propagation.}
    \label{fig:couette_cylinder_propagation}
\end{figure*}

The analytical solution predicts a $1/r$ decay of the displacement field near the inner cylinder. Therefore, the strain scales as $1/r^2$ and the strain energy as $1/r^4$. This produces a pronounced localization of the prescribed deformation: large strains (with $||\varepsilon|| \sim 1$) develop in a narrow region close to the inner boundary, while the deformation remains smooth and azimuthally homogeneous. As a result, the strain energy is uniform in the angular direction but sharply concentrated near the inner cylinder, where it quickly exceeds the fracture threshold and initiates crack formation.
Fracture develops around the inner cylinder, as shown in Figure~\ref{fig:couette_cylinder_propagation}. The fracturing pattern is symmetric, reflecting the rotational symmetry of the geometry together with small numerical perturbations due to finite resolution. Since the initial state is homogeneous, there is no preferred angular position for a localized fracture nucleation. In this configuration, the strain energy reaches the fracture threshold virtually uniformly along the inner boundary. As a result, damage occurs spontaneously all around the inner boundary, such that the fractured zone forms a continuous circumferential band rather than a propagating crack. This is in contrast to the uniaxial tensile case and simple plane shear cases, where the deformation is initially quite homogeneous but geometric asymmetry introduced by the inclusion and localized stress concentration lead to a single dominant fracture path with a well-defined crack tip.

Here, the imposed rotation generates an approximately axi-symmetric shear stress field, producing strong radial localization of strain energy while maintaining angular uniformity. The response in this idealized Couette geometry is therefore a narrow annular damage zone, where fracture is spatially distributed rather than tip-driven. This behavior is consistent with shear-band–like localization observed in sea-ice, where deformation organizes into narrow failure zones.
\section{Discussion and conclusions}\label{sec5}
We propose a minimal phase-field framework for fracture propagation, motivated by future applications to sea-ice, as a strongly forced geophysical system in which fracture plays a central role in controlling large-scale motion and regulating exchanges of heat, gas, and momentum between ocean and atmosphere \cite{kwok2001, serreze2011, ipcc_underestimation, Herman2012, goosse2018, Taylor2018}. The model has been validated across a set of standard benchmark loading configurations.

Under tensile loading, the model reproduces fracture mechanics within linear elasticity. Cracks propagate perpendicular to the maximum tensile stress, and their propagation speed scales with the elastic energy excess above the fracture threshold\cite{Karma_Kessler_Levine, anderson2005}. Under simple, plane shear loading, cracks develop along the principal stress directions \cite{anderson2005}, forming symmetric branching patterns, with propagation speeds again proportional to the applied load close to the critical threshold of fracturing energy. These results demonstrate that the model captures well the branching process and fracture propagation velocity without explicit tracking of crack surfaces. In the Couette geometry~\cite{cohesion2016, floating_ice_plate}, the localized strain energy along the inner cylinder leads to simultaneous fracture initiation rather than a propagating front, highlighting the role of energy distribution in determining fracture patterns. This contrasts with isolated stress concentrations, which favor directional crack propagation, and provides insight into shear band formation in sea-ice under distributed versus localized forcing. Overall, our phase-field approach offers a versatile and computationally robust tool for modeling fracturing under an applied body force and in damped dynamic conditions.

The choice of a double-well free-energy formulation \cite{Karma_Kessler_Levine}, in contrast to the more commonly employed Francfort–Marigo framework in phase-field fracture modeling \cite{francfort_marigo_1998, ambrosio_tortorelli_1990}, is motivated by the need to retain the overdamped fracture dynamics (associated with friction along rough interfaces) and to account for the coexistence of two phases of the material (intact ice and granular ice), rather than a clear fracture being the interface itself within a single intact stable phase. This is particularly relevant for sea-ice, where loading can vary rapidly and the typical time scales of interest are shorter than one day and for which our objective is to capture within the model the transition from intact to granular ice within shear bands. The numerical implementation based on spectral methods in Fourier space is well suited both to the strong nonlinearities introduced by the double-well free-energy functional and to the type of loading considered here, where body forces consistent with geophysical forcings are prescribed instead of classical Dirichlet boundary conditions.

Very few studies have been conducted on modeling fractures in sea-ice or ice at the geophysical scale using a phase-field approach. \cite{sondershaus2023} developed a phase-field fracture model for ice shelves that explicitly accounts for the visco-elastic behavior of ice. In particular, they employed a Maxwell-type visco-elastic rheology, which is highly relevant for future model developments based on the present study. Their variational framework targets loading scenarios that differ substantially from those considered here. \cite{giannakis2022} proposed a variational phase-field framework to describe brittle fracture in sea-ice, focusing on how stochastic ice thickness heterogeneities influence fracture strength and orientation under boundary-driven loading. Their approach is well-suited to capture mode-I fracture initiation and propagation in heterogeneous media. Finally, \cite{fei_choo2} investigated shear fractures in geological materials using a phase-field formulation that is, in spirit, much closer to our work. Because friction along discontinuities plays a central role in shear-fracture dynamics, their model introduced a detailed description of frictional interfaces and incorporated frictional dissipation directly into the energy balance. While this is highly relevant and sophisticated, their research philosophy diverges from ours in the complexity and richness of the constitutive ingredients employed and in that they worked under a quasi-static assumption, whereas we use here a quasi-dynamic representation of the fracture and damage processes.

Our work differs from these three approaches both in the physical regime targeted and in the modeling philosophy. Rather than discrete mode-I fractures \cite{giannakis2022} or detailed visco-elastic and frictional processes \cite{sondershaus2023,fei_choo2}, we aim to describe the granularization transition within sea-ice under compressive shear loading, focusing on the emergent geometry of fracture patterns, and, to this end, we prioritize capturing the phenomenology of the fracturing process over adhering to a strictly variational convergence framework. Moreover, whereas the above studies relied on boundary forcings and imposed displacements, we consider body forcing as the primary driver. Finally, we deliberately pursue a model with as few ingredients as possible, constructed to be as simple and minimalist as is consistent with the target physics. In line with this philosophy, our formulation is more closely related to a KKL-type framework, designed to capture continuous, quasi-dynamic damage accumulation, rather than the detailed propagation of single or few discrete cracks. Our results suggest that minimal, dynamically consistent phase‑field models can successfully capture features of sea‑ice shear fracturing and provide a practical foundation on which more detailed rheologies and geophysical forcings can be built.

The present study paves the way for future work that will focus on incorporating tension–compression asymmetry \cite{henry_levine2004,spatschek2010}, different material rheologies \cite{dansereau2016}, and realistic forcing to better represent the mechanical behavior of sea-ice and advancing our understanding of the granular transition within a shear band. Such insights could help determine whether the floes generated in such shear bands exhibit a characteristic length scale in their size distribution, and to quantify how these localized fracture events modify the statistical mechanical properties of sea-ice.
%\backmatter
\bmsection*{Author contributions}
LD performed the research and wrote the initial draft. VD designed and funded the research. VD, VS, LA, FR contributed to the design of the research, the writing, and supervision. 
\bmsection*{Use of generative AI}
During the preparation of this work, generative AI tools, such as ChatGPT and GPT UiO (\url{https://gpt.uio.no}), were used for language refinement and sentence rephrasing. All scientific content, research design, data analysis, results and conclusion were conducted solely by the authors. The authors reviewed all edits and take full responsibility for the final manuscript.
\bmsection*{Acknowledgments}
The authors thank Andres Leon Baldelli, Hervé Henry and Juan Michael Sagardo for helpful discussions and feedback on modeling choices during the development of this study. LD and VD acknowledge financial support from Schmidt Sciences through the SASIP project and the associated scientific community. We also thank Olivier Gagliardini and Joachim Mathiesen for their constructive comments throughout this work.
\bmsection*{Financial disclosure}
This research received support from Schmidt Sciences, LLC, via the SASIP project (grant G-24-67788).
\bmsection*{Conflict of interest}
The authors declare no conflict of interest.
\bmsection*{Data availability}
Source code and data used in this study are available from the corresponding author upon reasonable request.
\bibliography{wileyNJD-AMA}
\end{document}